\documentclass[aps,pra,twocolumn,superscriptaddress]{revtex4-1}

\usepackage[T1]{fontenc}
\usepackage[latin9]{inputenc}
\usepackage{amsmath}
\usepackage{amssymb}
\usepackage{graphicx}
\usepackage{hyperref}
\usepackage{natbib}
\usepackage{esint}
\usepackage{times}
\usepackage{helvet}
\usepackage{braket}
\usepackage{bbm}
\usepackage{xcolor}
\usepackage{enumitem}
\usepackage[toc,page,titletoc]{appendix}

\makeatletter
\newcommand{\JILA}{JILA, NIST and Department of Physics, University of Colorado, Boulder, CO 80309, USA}
\newcommand{\CTQM}{Center for Theory of Quantum Matter, University of Colorado, Boulder, CO 80309, USA}
\newcommand{\Physics}{Department of Physics, University of Colorado, Boulder, CO 80309}

\begin{document}
\title{Tunable-spin-model generation with spin-orbit-coupled fermions in optical lattices}
\date{\today}
\author{Mikhail Mamaev}
\email{mikhail.mamaev@colorado.edu}
\affiliation{\JILA}
\affiliation{\CTQM}
\author{Itamar Kimchi}
\affiliation{\JILA}
\affiliation{\CTQM}
\affiliation{\Physics}
\author{Rahul M. Nandkishore}
\affiliation{\CTQM}
\affiliation{\Physics}
\author{Ana Maria Rey}
\affiliation{\JILA}
\affiliation{\CTQM}

\begin{abstract}We study the dynamical behaviour of ultracold fermionic atoms loaded into an optical lattice under the presence of an effective magnetic flux, induced by spin-orbit coupled laser driving. At half filling, the resulting system can emulate a variety of iconic spin-1/2 models such as an Ising model, an XY model, a generic XXZ model with arbitrary anisotropy, or a collective one-axis twisting model. The validity of these different spin models is examined across the parameter space of flux and driving strength. In addition, there is a parameter regime where the system exhibits chiral, persistent features in the long-time dynamics. We explore these properties and discuss the role played by the system's symmetries. We also discuss experimentally-viable implementations.
\end{abstract}

\maketitle

%%%%%
\section{Introduction}
%%%%%
Understanding and quantifying the behaviour of interacting quantum particles in lattices is a fundamental goal of modern quantum science. While there is a plethora of research directions, one vital aspect is the response of particles to externally-imposed magnetic fields. Such fields induce an effective flux that threads through the plaquettes of the lattice~\cite{dalibard2011colloquium,goldman2014colloquium}, coupling the charge and spin degrees of freedom and modifying the particle dynamics. Interpreting the dynamical response to an applied flux is important for many applications in condensed matter, including ferroelectrics~\cite{Scott954}, spintronics~\cite{duine2018DMSpintronics,Hirohata} and spin-glass physics~\cite{fert1980DMSpinGlass,Goremychkin}.

One of the best ways to study such phenomena is with ultracold atomic experiments. State-of-the-art ultracold systems provide exceptional levels of cleanliness, isolation and tunability. They allow for pristine implementations of iconic Fermi- or Bose-Hubbard models that describe interacting particles in a lattice, with additional terms to account for the synthetic magnetic fields~\cite{Yoshiro2020}. A magnetic flux is easy to impose and control with tools such as laser driving, using Raman couplings or direct optical transitions.

There has been a great deal of theoretical work on Fermi- or Bose-Hubbard models with synthetic gauge fields that ultracold experiments could investigate, exploring ground-state phases~\cite{inbook,Aidelsburger,pixley2017strongCoupling} or phenomena such as many-body localization~\cite{suthar2020mbl}. However, the interplay of magnetic flux together with particle interactions can lead to complex dynamical behaviour that still lacks a good theoretical understanding. In the case of fermions, even non-interacting atoms can exhibit non-trivial behaviour due to Pauli exclusion, while the addition of Hubbard repulsion renders the dynamics even more complex.

In this work, we focus on fermionic atoms loaded into 3D optical lattices. In the Mott insulating limit with one atom per site, each atom acts as an effective spin and the physics can be simplified by mapping to an effective spin model. This spin model emerges via virtual atomic tunneling processes that lead to second-order superexchange interactions between these spins. Conventional lattice systems are often captured by isotropic Heisenberg models, while the presence of magnetic flux allows for tunability of the spin interactions. As we show here, the flux leads to more elaborate spin models such as anisotropic XXZ models, collective one-axis twisting Hamiltonians~\cite{Kitagawa1993}, or Dzyaloshinskii-Moriya (DM) interactions~\cite{dzyaloshinsky1958DM,moriya1960DM}. The latter is particularly intriguing, as it exhibits chirality, topological features and complex phase diagrams depending on the flux~\cite{peotta2014xyzBosonsSOCDMPhaseDiagram, garate2010DMphaseDiagram}. A system that can realize several different spin models with easy tunability can be very useful to the field, as current-generation optical lattice experiments have only recently begun to probe anisotropic interacting spin physics~\cite{jepsen2020spinTransportKetterle,Dimitrova2020}.

Here we study the case when the internal atomic states are driven by an external laser, which imprints a site-dependent phase that emulates a magnetic flux. We use the drive strength and the magnetic flux as tuning parameters, and show the different types of spin interactions that can be realized. We explore the dynamical properties of these interactions and specify regimes in the parameter space of flux and driving strength where the system's time evolution can be captured by simple models such as Ising, XY, XXZ or collective-spin one-axis twisting. We also study a regime where the corresponding spin model maps to a Heisenberg model in a twisted frame, causing the long-time dynamics to develop non-trivial features such as infinite-time magnetization and chiral spin imbalance. These latter features are not limited to the strongly interacting regime, but also hold in the weakly interacting limit of the Fermi-Hubbard model where atomic motion is relevant. We discuss the role that symmetry %and resulting Hilbert space fragmentation 
%RN: fragmentation is generally used to mean the Hilbert space breaks up even within symmetry sectors. I wouldn't call breakup into symmetry sectors fragmentation. 
plays in preventing the system from relaxing. Our predictions can be readily implemented in many ultracold systems, and are especially relevant for alkaline-earth or earth-like atoms in 3D lattices or tweezer arrays, which provide exceptional coherence times~\cite{Hutson2019,Norcia,young2020tweezer} to overcome the inherent slow interaction rates while avoiding issues from heating given the very low spontaneous emission rates of their low-lying electronic levels. 

The outline of the paper is as follows. Section~\ref{sec_Models} introduces the underlying Fermi-Hubbard model and derives the corresponding effective spin models that emerge at half filling. A dynamical classification of the spin model behaviour in different parameter regimes is given. Section~\ref{sec_DM} focuses on the low drive regime, and discusses the persistent magnetization or chiral features that can be observed due to the additional symmetry of a Heisenberg model in a twisted frame. Section~\ref{sec_Experiment} provides a detailed discussion on possible experimental implementations.

%probes of dynamics in XY + DM interaction ~\cite{derzhko2006dynamicProbesDMwithXY}
%%%%%
\section{Tunable spin dynamics}
\label{sec_Models}
%%%%%
%%%
\subsection{Fermi-Hubbard and spin model}
\begin{figure}[htb]
\centering
\includegraphics[width=1\linewidth]{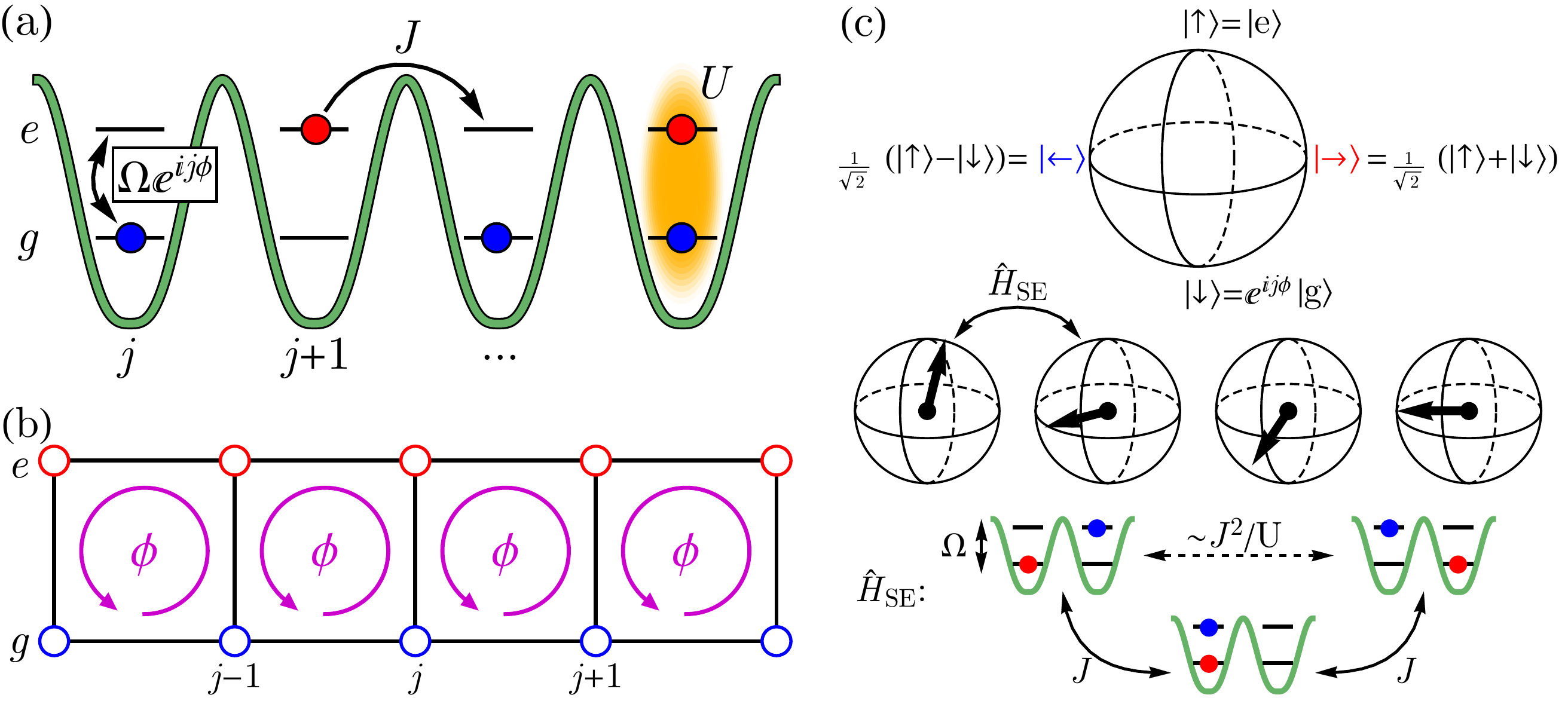}
\caption{(a) Schematic of Fermi-Hubbard dynamics in an optical lattice, with nearest-neighbour tunneling rate $J$, onsite repulsion $U$, and on-site driving $\Omega$. When the laser drive has a wavelength that is incommensurate with the underlying optical lattice wavelength, the laser   induces spin-orbit coupling (SOC) through a spatially-dependent phase $e^{i j \phi}$ in the drive term, with $\phi$ controlled by the drive implementation (e.g. laser alignment or lattice spacing, see Section~\ref{sec_Experiment}). (b) The SOC phase creates effective magnetic flux on plaquettes of the ladder state structure in 1D, with lattice index $j$ along the length of the ladder and internal states $e$, $g$ corresponding to the individual rungs. An atom tunneling around one plaquette picks up a total phase of $\phi$. (c) At half filling and strong interactions $U/J \gg 1$, the atomic spin $\sigma$ at different lattice site can be dressed by the laser drive, which modifies the resulting spin dynamics dominated by second-order virtual superexchange processes with rate $\sim J^2 / U$, up to possible normalization from the drive $\Omega$.}
\label{fig_Schematic}
\end{figure}

%%%
The system we describe is a three-dimensional (3D) optical lattice loaded with ultracold fermionic atoms cooled into the lowest motional band. Each atom has two internal states $\sigma \in \{g,e\}$ corresponding to a spin-1/2 degree of freedom, such as nuclear-spin polarized clock states split by an optical frequency, or two different nuclear-spin states within the ground hyperfine manifold. The effective system dimensionality is freely tuned by changing the lattice confinement strengths. We focus on the case where the system is effectively 1D by considering a strong confinement in two directions that suppresses tunneling along them. Along the direction atoms can tunnel, we assume a lattice of $L$ sites populated by $N$ atoms. The system is depicted in Fig.~\ref{fig_Schematic}(a). The Hamiltonian of the system is,
\begin{equation}
\begin{aligned}
\label{eq_FermiHamiltonian}
\hat{H}&=\hat{H}_{\mathrm{FH}}+\hat{H}_{\Omega}\\
\hat{H}_{\mathrm{FH}}&=-J\sum_{j,\sigma}\left(\hat{c}_{j,\sigma}^{\dagger}\hat{c}_{j+1,\sigma}+h.c.\right)+U\sum_{j}\hat{n}_{j,e}\hat{n}_{j,g}\\
\hat{H}_{\Omega}&=\frac{\Omega}{2}\sum_{j}(e^{i j \phi}\hat{c}_{j,e}^{\dagger}\hat{c}_{j,g}+h.c.).
\end{aligned}
\end{equation}
Here $\hat{c}_{j,\sigma}$ annihilates an atom with spin $\sigma$ on site $j$. Atoms tunnel at rate $J$ and exhibit onsite repulsion of strength $U$, proportional to the $a_{eg}^{-}$ singlet scattering length. We have assumed equal tunneling rates for both $g$ and $e$, which implies a magic-wavelength lattice if the internal states are clock states. In addition to the standard Fermi-Hubbard term $\hat{H}_{\mathrm{FH}}$, we include a laser-driving term $\hat{H}_{\Omega}$ that induces the desired flux $\phi$ through a spatially-dependent phase $e^{i j \phi}$, providing model tunability. The differential phase imprinted by the laser implements a net  spin-orbit coupling (SOC) by generating a momentum kick to the atoms while flipping their spin~\cite{dalibard2011colloquium,goldman2014colloquium,Wall2016,kolkowitz2017soc,Bromley2018,Livi2016}. The realization of this effective flux in 1D is depicted in Fig.~\ref{fig_Schematic}(b). The drive can be implemented with a direct interrogating laser (if $g$, $e$ are clock states), or a Raman coupling (if $g$, $e$ are ground hyperfine levels), see Section~\ref{sec_Experiment} for details. We assume $\phi \in [0,\pi]$ without loss of generality.

For sufficiently strong interactions $U/J \gg 1$ at half filling $N/L = 1$ with one atom per site, double occupancies of lattice sites are strongly suppressed. The system dynamics in this regime may be approximated with a spin model, as depicted in Fig.~\ref{fig_Schematic}(c). We define a dressed spin basis at different lattice sites by making a gauge transformation, defining new fermionic operators $\{\hat{a}_{j,\uparrow}, \hat{a}_{j,\downarrow}\}$ to remove the SOC phase,
\begin{equation}
\begin{aligned}
\label{eq_BasisRotation}
\hat{c}_{j,e}^{\dagger}\ket{0}&\equiv \hat{a}_{j,\uparrow}^{\dagger}\ket{0}\leftrightarrow\ket{\uparrow}_{j},\\
e^{-i j \phi}\hat{c}_{j,g}^{\dagger}\ket{0}&\equiv \hat{a}_{j,\downarrow}^{\dagger}\ket{0}\leftrightarrow\ket{\downarrow}_{j},
\end{aligned}
\end{equation}
and define conventional spin operators for these dressed atoms,
\begin{equation}
\begin{aligned}
\hat{\sigma}_{j}^{x}&=\left(\hat{a}_{j,\uparrow}^{\dagger}\hat{a}_{j,\downarrow}+h.c.\right),\\
\hat{\sigma}_{j}^{y}&=-i\left(\hat{a}_{j,\uparrow}^{\dagger}\hat{a}_{j,\downarrow}-h.c.\right),\\
\hat{\sigma}_{j}^{z}&=\hat{a}_{j,\uparrow}^{\dagger}\hat{a}_{j,\uparrow}-\hat{a}_{j,\downarrow}^{\dagger}\hat{a}_{j,\downarrow}.
\end{aligned}
\end{equation}
This is just the Abrikosov pseudo-fermion representation~\cite{AA, Coleman}. Standard second-order perturbation theory then leads to the following general superexchange (SE) spin model (see Appendix~\ref{app_SpinDerivation} for derivation),
\begin{equation}
\begin{aligned}
\label{eq_SpinModel}
\hat{H}_{\mathrm{SE}}&\approx  J_{\parallel}\sum_{j}\left(\hat{\sigma}_{j}^{x}\hat{\sigma}_{j+1}^{x}+\hat{\sigma}_{j}^{y}\hat{\sigma}_{j+1}^{y}\right)+J_{\perp}\sum_{j}\hat{\sigma}_{j}^{z}\hat{\sigma}_{j+1}^{z}\\
&+J_{\mathrm{DM}} \sum_{j}\left(\hat{\sigma}_{j}^{x}\hat{\sigma}_{j+1}^{y}-\hat{\sigma}_{j}^{y}\hat{\sigma}_{j+1}^{x}\right)+J_{\Omega}\sum_{j}\hat{\sigma}_{j}^{x}.
\end{aligned}
\end{equation}
The first two terms correspond to an XXZ model. The third term is a DM interaction with a plane axis of $\hat{z}$ [thus also taking the form of $\hat{z} \cdot (\vec{\sigma}_{j}\times \vec{\sigma}_{j+1})$]. The interaction coefficients are,
\begin{equation}
\begin{aligned}
\label{eq_SpinModelCoefficients}
J_{\parallel}&=\frac{J^2}{U}\frac{U^2\cos(\phi)-\Omega^2 \cos^2(\frac{\phi}{2})}{U^2-\Omega^2},\\
J_{\perp} &= \frac{J^2}{U}\frac{U^2-\Omega^2 \cos^2(\frac{\phi}{2})}{U^2-\Omega^2},\\
J_{\mathrm{DM}}&=\frac{J^2}{U}\frac{(\Omega^2-2U^2)\sin(\phi)}{2(U^2-\Omega^2)},\\
J_{\mathrm{\Omega}}&=\frac{\Omega}{2}-\frac{2J^2 \Omega \sin^2(\frac{\phi}{2})}{U^2-\Omega^2}.
\end{aligned}
\end{equation}
These share the conventional $J^2/U$ superexchange energy scale, with different normalization factors coming from the interplay between the drive and flux. This model is valid in the Mott insulating limit $U/J \gg 1$, for all flux values $\phi$, and all drive frequencies $\Omega$ far from the resonance point $|\Omega| = |U|$, requiring a spacing of $||U|-|\Omega||\gg J$ to prevent higher-order effects. Table~\ref{table1} shows some experimentally-realistic parameter values for which the model is expected to be valid, for the case of nuclear spin polarized  ultracold  $^{87}Sr$ atoms using their $^{1}S_{0}$, $^{3}P_{0}$ clock states as the internal states $e$, $g$. The lattice depth $(V_x,V_y,V_z)/E_r$ (with $E_r$ the recoil energy), tunneling rate $J$ and Hubbard repulsion $U$ are provided. Typical laser drive parameters of Rabi frequency and flux $(\Omega,\phi)$ are also given for some specific regimes of the general spin model above, which will be discussed in the following Section~\ref{sec_SpinModelRegimes}. There are two sets of sample parameters; the first has a very large $U/J \approx 50$, which we will use for model benchmarking later in Section~\ref{subsec_Benchmarking} to avoid additional error from not being in the Mott insulating limit $U/J \gg 1$. The second set offers more experimentally-realistic parameters. This case will have additional corrections from  a moderate $U/J$ value, but still at a level low enough to clearly observe the desired superexchange dynamics in an experimental setting.

Note that there can also be other resonances such as $|\Omega| \neq |U|/2$, as considered in e.g. Ref.~\cite{meinert2014resonantTilt}, though the relevant width is far smaller as will be seen in the next section. These spin interactions are anisotropic even if the drive is turned off, $\Omega = 0$, because we are using a dressed basis to absorb the SOC phase (see Appendix~\ref{app_SpinDerivation}). In the context of the underlying Fermi-Hubbard model, one can use a fast pulse of a SOC drive to prepare a desired initial state such as a product state in the dressed basis, after which the drive may be turned off if desired (see Section~\ref{sec_Experiment} for details). There is also a single-particle term $J_{\Omega}$ corresponding to the drive. This term contains an additional single-particle superexchange contribution, but this is typically negligible compared to the bare drive, and so we can approximate $J_{\Omega}\approx \Omega/2$.

\begin{table}[]
\centering
\caption{\label{table1}Table of experimentally-realistic parameters, using ultracold alkaline earth $^{87}$Sr with its $^{1}S_0$, $^{3}P_0$ clock states as $e$, $g$. The second column gives values that correspond to $U/J \approx 50$, matching the benchmarking in Section~\ref{subsec_Benchmarking}. The third column gives a set of values with a smaller $U/J \approx 25$ and thus faster dynamics.}
\begin{tabular}{|c|c|c|}
\hline
\textbf{Parameter} & \textbf{Value} & \textbf{Value (faster)} \\ \hline
$(V_x, V_y, V_z)/E_r$ & (13,100,100) & (10,100,100) \\ \hline
$J$ & 34 Hz & 66 Hz\\ \hline
$U$ & 1.8 kHz & 1.7 kHz \\ \hline
($\Omega,\phi$) & (1.5 kHz,$\pi$) & (1.1 kHz,$\pi$)\\
for Ising & $J_{\parallel}$=-2 Hz & $J_{\parallel}$=-5 Hz\\ \hline
($\Omega,\phi$) & (1.5 kHz,1.0) & (1.1 kHz,1.3) \\
for XY & $J_{\perp}/2$=0.5 Hz & $J_{\perp}/2$=2 Hz\\ \hline
($\Omega,\phi$) & (1.5 kHz,0.3) & (1.1 kHz,0.3)\\
for OAT & $2(J_{\parallel}-J_{\perp})$=-0.2 Hz & $2(J_{\parallel}-J_{\perp})$=-0.4 Hz\\ \hline
\end{tabular}
\end{table}

If the drive is off, $\Omega = 0$, this spin model commutes with and thus conserves total $\langle \hat{S}^{z}\rangle$, where $\hat{S}^{\gamma} = \frac{1}{2}\sum_{j}\hat{\sigma}_{j}^{\gamma}$. When the drive is instead very strong, $\Omega \gg J_{\parallel}, J_{\perp}, J_{\mathrm{DM}}$, total $\langle \hat{S}^{x}\rangle$ is approximately conserved instead because the drive imposes an energy penalty to flipping spins along the $\pm x$ Bloch sphere direction. Note that while the superexchange coefficients can be modified by the drive, as a first rough estimate the high-drive condition $\Omega \gg J_{\parallel}, J_{\perp}, J_{\mathrm{DM}}$ may be interpreted as a drive faster than the bare superexchange rate, $\Omega \gg J^2/U$.

%%%
\subsection{Spin model regimes}
\label{sec_SpinModelRegimes}
%%%
While the general spin model of Eq.~\eqref{eq_SpinModel} is complex, there are parameter regimes where its form simplifies, allowing the dynamical emulation of other more conventional spin models. For a strong drive $\Omega \gg J_{\mathrm{DM}}$, the DM interaction is averaged out and can be neglected. Furthermore, assuming the drive also satisfies $\Omega \gg J_{\parallel}, J_{\perp}$, the $\hat{\sigma}_{j}^{y}\hat{\sigma}_{j+1}^{y}$ and $\hat{\sigma}_{j}^{z}\hat{\sigma}_{j+1}^{z}$ terms are equivalent in unitary evolution under a rotating-wave approximation, allowing us to interchange and collect them together via $\hat{\sigma}_{j}^{y}\hat{\sigma}_{j+1}^{y} \approx \hat{\sigma}_{j}^{z}\hat{\sigma}_{j+1}^{z} \approx \frac{1}{2}(\hat{\sigma}_{j}^{y}\hat{\sigma}_{j+1}^{y} + \hat{\sigma}_{j}^{z}\hat{\sigma}_{j+1}^{z})$. This leaves an XXZ-type model,
\begin{equation}
\begin{aligned}
\label{eq_SpinModelXXZ}
\hat{H}_{\mathrm{XXZ}}&=J_{\parallel}\sum_{j}\hat{\sigma}_{j}^{x}\hat{\sigma}_{j+1}^{x}+\frac{J_{\parallel}+J_{\perp}}{2}\sum_{j}\left(\hat{\sigma}_{j}^{y}\hat{\sigma}_{j+1}^{y}+\hat{\sigma}_{j}^{z}\hat{\sigma}_{j+1}^{z}\right)\\
&+J_{\Omega}\sum_{j}\hat{\sigma}_{j}^{x}\>\>\>\>\>\>\>\>\>\>\>\>\>\>\>\>\>\>\>\>\>\>\>\>\>\>\>\>\>\>\>\>\>\>\>\>\text{[$\Omega \gg J_{\parallel},J_{\perp},J_{\mathrm{DM}}$]}.
\end{aligned}
\end{equation}
Note that this model is not the same as simply taking the XXZ-like piece from the first line of Eq.~\eqref{eq_SpinModel}, as here the single-particle drive term commutes with the XXZ term, making it easy to account for in unitary evolution. One could also write an XXZ model for the no-drive limit $\Omega = 0$ (see Appendix~\ref{app_SpinDynamics}), for which we would just keep the XXZ portion of Eq.~\eqref{eq_SpinModel}; this can be valid in the no-drive limit if the $J_{\mathrm{DM}}$ term vanishes parametrically due to its $\sin(\phi)$ factor for $\phi \approx 0, \pi$.

As one limiting regime of the XXZ model, in the strong-drive regime $\Omega \gg J_{\parallel}, J_{\perp}$,$J_{\mathrm{DM}}$ if we also have $\phi \approx \pi$, the coefficients $J_{\parallel}$ and $J_{\perp}$ are approximately equal and opposite ($J_{\parallel} \approx -J_{\perp}$), causing them to cancel each other out and leave an Ising model,
\begin{equation}
\begin{aligned}
\hat{H}_{\mathrm{Ising}}=J_{\parallel}\sum_{j}\hat{\sigma}_{j}^{x}\hat{\sigma}_{j+1}^{x}+J_{\Omega}&\sum_{j}\hat{\sigma}_{j}^{x}\\
&\text{[$\phi \approx \pi$, $\Omega \gg J_{\parallel}, J_{\perp}$,$J_{\mathrm{DM}}$]}.
\end{aligned}
\end{equation}

As another special regime of the XXZ model, there is a line in parameter space where the $J_{\parallel}$ coefficient vanishes, requiring $\Omega = \pm U \sqrt{\cos(\phi)} \sec(\phi/2)$, which causes the $(\hat{\sigma}_{j}^{x}\hat{\sigma}_{j+1}^{x}+\hat{\sigma}_{j}^{y}\hat{\sigma}_{j+1}^{y})$ terms of Eq.~\eqref{eq_SpinModel} to vanish. Assuming that we also still have a strong drive $\Omega \gg J_{\parallel}, J_{\perp}$,$J_{\mathrm{DM}}$, the DM interaction remains averaged out as well, leaving only the $\hat{\sigma}_{j}^{z}\hat{\sigma}_{j+1}^{z}$ term along with the drive (which is now transverse to the interaction). If we again make the rotating-wave approximation $\hat{\sigma}_{j}^{z}\hat{\sigma}_{j+1}^{z} \approx \frac{1}{2}(\hat{\sigma}_{j}^{y}\hat{\sigma}_{j+1}^{y} + \hat{\sigma}_{j}^{z}\hat{\sigma}_{j+1}^{z})$, we arrive at an XY model,
\begin{equation}
\begin{aligned}
\hat{H}_{\mathrm{XY}}=\frac{J_{\perp}}{2}&\sum_{j}\left(\hat{\sigma}_{j}^{y} \hat{\sigma}_{j+1}^{y}+\hat{\sigma}_{j}^{z}\hat{\sigma}_{j+1}^{z}\right)+\frac{\Omega}{2}\sum_{j}\hat{\sigma}_{j}^{x}\\
&\>\>\>\>\>\>\>\>\>\>\>\text{[}\Omega = U \sqrt{\cos(\phi)}\sec(\phi/2),\>\>\>\Omega \gg J_{\mathrm{DM}}\text{]}.
\end{aligned}
\end{equation}
We emphasize that this XY model with a strong (commuting) field, and the underlying Ising model with a strong transverse field are only equivalent under unitary time-evolution~\cite{kiely2018isingAndXY}.

For the strong drive regime $\Omega \gg J_{\parallel}, J_{\perp}$,$J_{\mathrm{DM}}$ with small flux, the coefficients $J_{\parallel}$, $J_{\perp}$ are almost equal to $J^2/U$. We can thus collect the XXZ model of Eq.~\eqref{eq_SpinModelXXZ} into an isotropic Heisenberg term and a perturbative $\hat{\sigma}_{j}^{x}\hat{\sigma}_{j+1}^{x}$ component. In this regime the system can be approximated with a collective-spin one-axis twisting (OAT) model, whose dynamical properties can be explored using restrictions to the fully-symmetric Dicke manifold~\cite{rey2008gapProtection,cappellaro2009gapProtection,kwasigroch2014gapProtection,perlin2020squeezing}. The model is written as,
\begin{equation}
\begin{aligned}
\hat{H}_{\mathrm{OAT}}&=\hat{P}_{\mathrm{Dicke}}\hat{H}_{\mathrm{XXZ}}\hat{P}_{\mathrm{Dicke}}\\
&=\frac{2(J_{\parallel}+J_{\perp})}{L-1}\vec{S}\cdot \vec{S} + \frac{2(J_{\parallel} - J_{\perp})}{L-1}\hat{S}^{x}\hat{S}^{x} + 2J_{\Omega}\hat{S}^{x},\\
&\>\>\>\>\>\>\>\>\>\>\>\>\>\>\>\>\>\>\>\>\>\>\>\>\>\>\>\>\>\>\>\>\>\>\>\>\>\>\>\>\>\>\>\>\> [\phi L \ll  1, \Omega \gg J_{\parallel},J_{\perp},J_{\mathrm{DM}}].
\end{aligned}
\end{equation}
The operator $\hat{P}_{\mathrm{Dicke}}$ projects to the Dicke manifold, spanned by the collective-spin states $\ket{S=L/2,M}$ which are eigenstates of $\hat{S}^{x}\ket{S,M}=M\ket{S,M}$ and 
$\vec{S}\cdot \vec{S}\ket{S,M}=S(S+1)\ket{S,M}$, with $\vec{S} = (\hat{S}^{x}, \hat{S}^{y}, \hat{S}^{z})$. Here $M$ takes values  $-S,-S+1,\dots S-1,S$. The flux must be small compared to $1/L$ rather than to 1, because the collective regime validity depends on the Dicke manifold gap, which shrinks with system size for nearest-neighbour interactions. Note that while the above model requires a strong drive, like the XXZ model we can also write a one-axis twisting model in the $\Omega = 0$ regime as well (see Appendix~\ref{app_SpinDynamics}). Similar low-drive collective physics were explored in Ref.~\cite{he2019squeezing} for the weakly-interacting regime; in contrast, here we have a strongly-interacting model that nonetheless allows us to map the nearest-neighbour superexchange interactions to a collective-spin model through gap protection. Moreover, as discussed in Ref.~\cite{perlin2020squeezing}, the collective behaviour can be more robust when mapping from an XXZ model with anisotropy slightly below unity (on the easy-plane side), which this model can realize as discussed in the next section.

Finally, for a small drive $\Omega \lesssim J_{\parallel}, J_{\perp}$,$J_{\mathrm{DM}}$, we must use the full model of Eq.~\eqref{eq_SpinModel}. If the drive is turned off completely ($\Omega = 0$), the resulting interaction can be written as,
\begin{equation}
\label{eq_HeisenbergTwist}
\begin{aligned}
\hat{H}_{\mathrm{Heisen+T}} =\frac{J^2}{U}&\sum_{j}\big[\cos(\phi)\left(\hat{\sigma}_{j}^{x}\hat{\sigma}_{j+1}^{x}+\hat{\sigma}_{j}^{y}\hat{\sigma}_{j+1}^{y}\right)+\hat{\sigma}_{j}^{z}\hat{\sigma}_{j+1}^{z} \\
- &\sin(\phi)\left(\hat{\sigma}_{j}^{x}\hat{\sigma}_{j+1}^{y}-\hat{\sigma}_{j}^{y}\hat{\sigma}_{j+1}^{x}\right)\big],\>\>\>[\Omega = 0].
\end{aligned}
\end{equation}
We label the spin interaction in this regime as Heisenberg+twist (Heisen+T). While a DM interaction is present, its effect for the above parameters can be simplified to Heisenberg model physics in a twisted frame of reference, as will be discussed in Section~\ref{sec_DM}. Having a small non-zero $\Omega \lesssim J_{\mathrm{DM}}$ will not significantly change this picture aside from adding the corresponding single-particle term $J_{\Omega} \sum_{j}\hat{\sigma}_{j}^{x}$, since the coefficients of the model are only weakly dependent on $\Omega$ in this regime. For a larger $\Omega \gg J_{\mathrm{DM}}$, the DM term will be rotated out.

In addition to all of the above, we can also add an extra field through the use of the laser drive's detuning $\delta$, which would take the form of $\frac{\delta}{2}\sum_{j}(\hat{n}_{j,e}-\hat{n}_{j,g})$ in the basis of the bare Fermi-Hubbard model, thus adding a spin term of the form $\sim \frac{\delta}{2} \sum_{j}\hat{\sigma}_j^{z}$. While such a term does not commute with the rest of the Hamiltonian, if $|\delta| \ll |U^2 - \Omega^2|$, the superexchange model will remain the same to good approximation. This permits the addition of an extra single-particle term without changing the spin interactions. We do not explicitly do so in this work, but such a detuning nonetheless provides yet another tuning parameter that can be implemented without the need for additional experimental ingredients.

%%%
\subsection{Dynamical model comparisons}
\label{subsec_Benchmarking}
%%%
It is useful to know where the various simplified models discussed in the previous section are applicable. The most rigorous metric of dynamical model agreement is state fidelity, but such a comparison tends to be unnecessarily harsh because the fidelity can drop with increasing system size while experimentally-relevant observables remain in agreement. We instead evaluate the validity of the spin models through comparison of simple collective observables.

We examine two typical time evolutions of the system, starting from product initial states. The first evolution is,
\begin{equation}
\ket{\psi_0^{(Z)}}=\bigotimes_{j}\ket{\uparrow}_{j} \text{  measuring   } \mathcal{C}^{(Z)}=\frac{2}{L}\sqrt{\langle \hat{S}^{y}\rangle^2 +\langle \hat{S}^{z}\rangle^2}.
\end{equation}
The observable $\mathcal{C}^{(Z)}$ is the spin contrast, chosen such that there are no fast single-particle oscillations coming from the drive. The second evolution is,
\begin{equation}
\ket{\psi_0^{(X)}}=\bigotimes_{j}\ket{\rightarrow}_{j} \text{  measuring   } \mathcal{C}^{(X)}=\frac{2}{L}\langle \hat{S}^{x}\rangle,
\end{equation}
where $\ket{\rightarrow}_j = (\ket{\uparrow}_j + \ket{\downarrow}_j)/\sqrt{2}$. Here we label our contrast $\mathcal{C}^{(X)}$ as the magnetization. For a large drive $\Omega \gg J_{\parallel}, J_{\perp}, J_{\mathrm{DM}}$ only the first $\mathcal{C}^{(Z)}$ evolution will see non-trivial dynamics, as $\mathcal{C}^{(X)}$ is approximately conserved by the drive. On the other hand, for a weak drive $\Omega \lesssim J_{\parallel}, J_{\perp}, J_{\mathrm{DM}}$ we can have non-trivial dynamics for both evolutions depending on the model in question. In principle a full rigorous comparison should consider all possible high-energy state properties rather than the two selected above. Since we mainly seek a qualitative understanding of model regimes, we have chosen a pair of experimentally-simple evolutions for which at least one will have non-trivial dynamics at every point in parameter space.

To determine the validity of a particular spin model, we compute the time-dependence of $\mathcal{C}^{(\alpha)}$ for both evolutions $\alpha \in \{Z,X\}$ using a spin model $\mathcal{C}_{\mathrm{Spin}}^{(\alpha)}$ (where $\mathrm{Spin} \in \{\mathrm{Ising, XY, OAT, XXZ, Heisen+T}\}$) and the Fermi-Hubbard model $\mathcal{C}_{\mathrm{Fermi}}^{(\alpha)}$ [re-written to reflect the basis rotation of Eq.~\eqref{eq_BasisRotation}], out to a time of,
\begin{equation}
t_f = 4/\text{max}(|J_{\parallel}|, |J_{\perp}|),
\end{equation}
which is four times the timescale of the fastest superexchange interaction strength at any given point in parameter space. This timescale is used for every model except the OAT, for which we instead use $t_f = 8/|J_{\parallel}-J_{\perp}|$ (16 time units of the twisting term $(\hat{S}^{x})^2$ without the $L$-dependence, which is sufficient to observe entanglement properties such as spin-squeezing). A dynamical error metric for a given spin model is defined as,
\begin{equation}
\Delta^{(\alpha)}_{\mathrm{Spin}} = \sqrt{\frac{1}{t_f}\int_{0}^{t_f} dt \left[\mathcal{C}_{\mathrm{Fermi}}^{(\alpha)}(t) - \mathcal{C}_{\mathrm{Spin}}^{(\alpha)}(t)\right]^2}.
\end{equation}
This metric is a root-mean-square error giving the average difference between a contrast measurement of the Fermi and spin model being considered over the time interval $[0, t_f]$. For every choice of $\phi$, $\Omega$ in the parameter space there will be two error metrics for the two evolutions $\alpha \in \{Z,X\}$, and we take the worse of the two,
\begin{equation}
\label{eq_ErrorMetricFinal}
\Delta_{\mathrm{Spin}} = \mathrm{max}\left[\Delta^{(Z)}_{\mathrm{Spin}},\Delta^{(X)}_{\mathrm{Spin}}\right].
\end{equation}
Of course, the disagreement between the models will tend to grow at longer times, and to an extent this analysis is qualitative. We choose four times the superexhange rate because this should be sufficient to see non-trivial contrast decay, and be useful for applications such as spin-squeezing or entangled state generation.

\begin{figure*}[htb]
\centering
\includegraphics[width=1\linewidth]{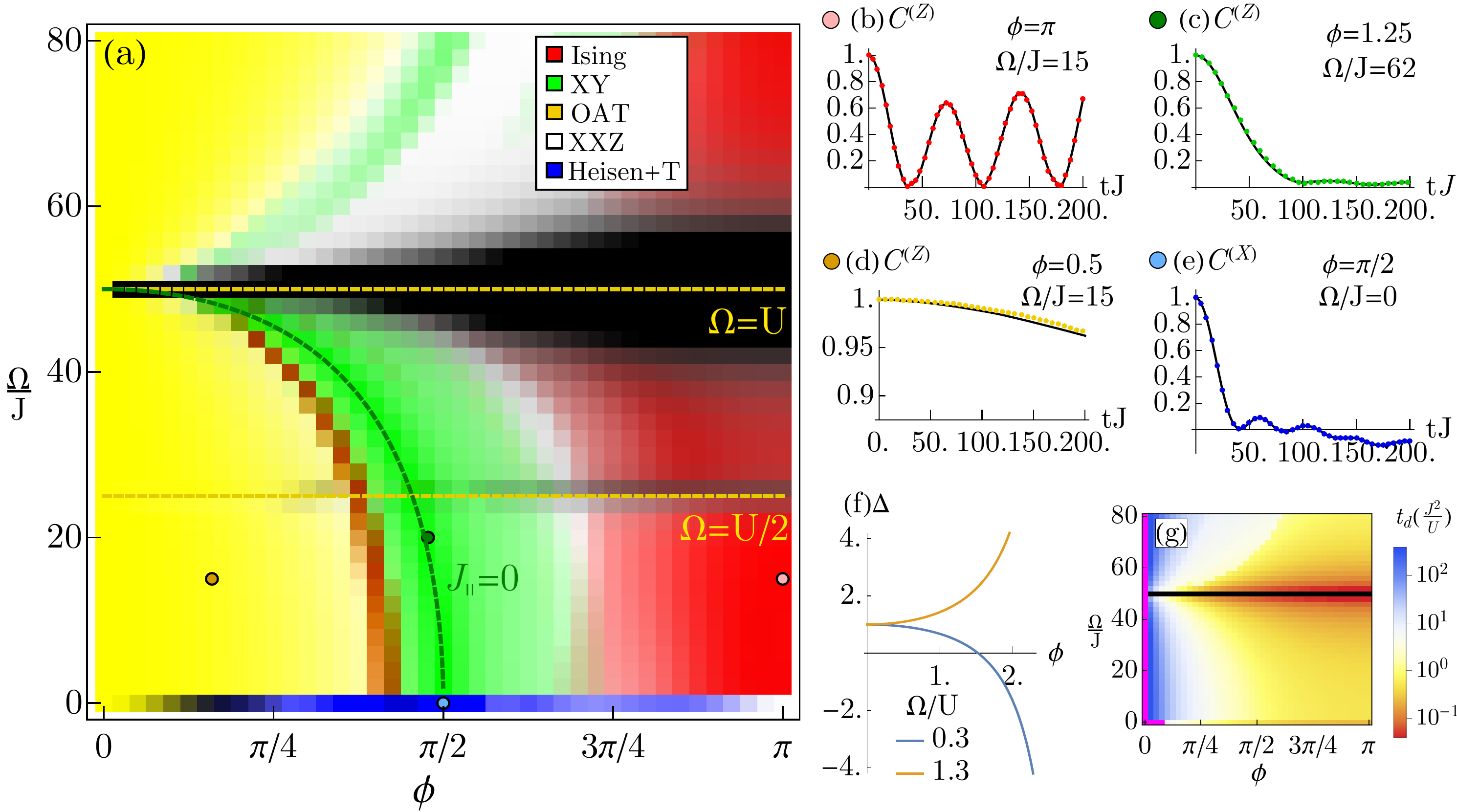}
\caption{(a) Dynamical regimes of the driven Fermi-Hubbard model compared to various spin models. The system (size $L=8$) is evolved to a fixed time $t_f$ (four times the fastest superexchange rate), and the contrast observable of the Fermi model is compared to each listed spin model through an error metric [Eq.~\eqref{eq_ErrorMetricFinal}]. The color scheme depicts regions where the error metric of the corresponding spin model is small; the color of a given model fully vanishes when $\Delta_{\mathrm{Spin}}\geq 0.25$, corresponding to an average error of 0.25 in a contrast measurement (see Appendix~\ref{app_SpinDynamics} for more details on the color scaling). Regions in black indicate regimes where the spin model description breaks down altogether due to higher-order processes from resonances such as $\Omega = U$ or $\Omega = U/2$, with the RGB color coordinates scaled down by the error of the full spin model $\Delta_{\mathrm{SE}}$. (b-e) Snapshots of contrast evolutions for the Ising, XY, OAT, XXZ and Heisen+T (Heisenberg+twist) models respectively, for specific points in the parameter regime as indicated by filled circles in panel (a). The former three plot the evolution of the high-drive contrast $\mathcal{C}^{(Z)}$, while the Heisen+T plot shows evolution of the low-drive contrast $\mathcal{C}^{(X)}$. (f) Anisotropy of the XXZ model as a function of flux for different fixed values of drive strength relative to Hubbard repulsion. (g) Characteristic timescale $t_d$ needed for the contrast to decay down to $1/e$, using $\mathcal{C}^{(Z)}$ for high drive $\Omega/J \gg J^2/U$ and $\mathcal{C}^{(X)}$ for no drive $\Omega = 0$. Black regions are points where $\Omega = U$, and the spin model description is invalid due to resonance. Purple points indicate parameters for which the contrast does not decay below $1/e$ at any time. This occurs trivially for $\phi  = 0$ (where the spin model is a pure Heisenberg model and no dynamics occur for product states), and for the special regime of $\phi \ll 1$, $\Omega = 0$ (which is discussed in Section~\ref{sec_DM}).}
\label{fig_PhaseDiagram}
\end{figure*}

Fig.~\ref{fig_PhaseDiagram}(a) shows a color plot of the resulting error metrics for the different spin models. The color scheme is chosen such that a given model's color completely vanishes when its error metric reaches $\Delta_{\mathrm{Spin}}\geq 0.25$, corresponding to an average error of 0.25 in measurements of $\mathcal{C}^{(X)}$ or $\mathcal{C}^{(Z)}$. The Hubbard interaction strength is chosen to be $U/J = 50$ as in the first sample set of parameters in Table~\ref{table1}, well into the Mott insulating regime to ensure that finite $U/J$ does not contribute additional error. We see the parameter regimes as described in Section~\ref{sec_SpinModelRegimes}. Large flux $\phi \approx \pi$ and strong drive corresponds to the Ising model (red). The line in parameter space satisfying $J_{\parallel} = 0$ corresponds to the XY model (green). There is a narrow red line adjacent to this regime where the Ising model also looks to be valid, but this is a spurious effect caused by the Ising evolution happening to align with the Fermi-Hubbard out to the specific timescale $t_f$ we use. For small flux $\phi L \ll 1$ and large drive, we have the OAT model (yellow). Underlying the other models is the XXZ model in white, which is valid throughout most of the parameter regime. The XXZ anisotropy parameter $\Delta = 2J_{\parallel} / (J_{\parallel}+J_{\perp})$ is plotted in Fig.~\ref{fig_PhaseDiagram}(f); we can attain any ferromagnetic or easy-plane value for $\Omega /U < 1$, and any antiferromagnetic value for $\Omega / U > 1$. Finally, for low drive $\Omega = 0$ we have the Heisenberg+twist model (blue). Note that the system in this regime can also be described by a low-drive version of the XXZ or OAT models (see Appendix~\ref{app_SpinDynamics}), which is why the corresponding colors are also present for $\Omega=0$. The regions in blue are points in parameter space where the chiral DM interaction exclusive to the Heisenberg+twist is necessary to capture the correct time-evolution.

Figs.~\ref{fig_PhaseDiagram}(b-e) show snapshots of the relevant models' time-evolution from different points of the diagram. The dark region near $\Omega = U$ is the resonance point where the superexchange denominators vanish and second-order perturbation theory breaks down, causing no spin model to be valid. There is an additional resonance point near $\Omega = U/2$ corresponding to a second-order resonant process not captured by the spin model~\cite{meinert2014resonantTilt}, although the width of this resonance is smaller. We also note that the associated timescales speed up as the resonance point $\Omega = U$ is approached. Fig.~\ref{fig_PhaseDiagram}(g) plots a characteristic timescale $t_d$ defined as the time needed for the contrast to decay down to $1/e$, using the $\mathcal{C}^{(Z)}$ contrast for $\Omega/J \gg J^2/U$ (the high drive limit, all data points except the $\Omega = 0$ line) and $\mathcal{C}^{(X)}$ for $\Omega = 0$. This allows for an evaluation of experimental tradeoff, where one can move closer to the resonance for faster timescales at the cost of weaker model agreement. There is also a region of $\phi \ll 1$, $\Omega = 0$ where the contrast does not fully decay at any time despite no obvious conservation law protecting it. This persistent magnetization effect will be discussed in Section~\ref{sec_DM}.

While these simulations are for relatively small system size $L=8$, in general the regimes of validity do not undergo significant change as the size increases because the interactions are nearest-neighbour. The only exception is the OAT model, for which the regime will shrink with increasing $L$ because it requires $\phi L \ll 1$ (the yellow region looks relatively large here because we use a small $L$). We also note that open boundary conditions are used for the above simulations with all models except the OAT; this is to ensure that the chiral properties of the DM interaction are captured, as will be explored in Section~\ref{sec_DM}. Comparisons with the OAT use a periodic Fermi-Hubbard model because open boundaries can lead to a minor but non-zero offset to the contrast even in the thermodynamic limit.

For the simpler regimes such as the Ising or one-axis twisting model, the dynamics are well understood and can have analytic solutions~\cite{foss2013IsingSolutions}. More general XXZ-type dynamics can be complex to treat, as even exact 1D Bethe ansatz techniques are difficult for full dynamical evolution. However, the parameter regime of low drive $\Omega = 0$ where the Heisenberg+twist model is valid offers a special case. The dynamics there are non-trivial, but exhibit special long-time features that can be understood from even the non-interacting limit, offering analytic tractability while still simulating the dynamical behaviour of a strongly interacting model. In the next section, we will focus on this regime in more detail.

%%%%%
\section{Persistent long-time behaviour}
\label{sec_DM}
%%%%%
%%%
\subsection{Long-time magnetization profiles}
%%%
Having shown the different regimes of spin models that can be realized with laser-driven SOC optical lattice systems, we focus on the regime of $\Omega = 0$ where the DM interaction plays a role. Conventionally, the ground-state properties of systems including DM interactions can already be quite complex~\cite{garate2010DMphaseDiagram}. In our case, we can work directly with the Heisenberg+twist model of Eq.~\eqref{eq_HeisenbergTwist}. For clarity, we write it again,
\begin{equation}
\begin{aligned}
\label{eq_SpinModelDM}
\hat{H}_{\mathrm{Heisen+T}} =\frac{J^2}{U}\cos(\phi)\sum_{j}\big[\hat{\sigma}_{j}^{x}&\hat{\sigma}_{j+1}^{x}+\hat{\sigma}_{j}^{y}\hat{\sigma}_{j+1}^{y}+\Delta_{\Omega=0}\hat{\sigma}_{j}^{z}\hat{\sigma}_{j+1}^{z} \\
&+ D\left(\hat{\sigma}_{j}^{x}\hat{\sigma}_{j+1}^{y}-\hat{\sigma}_{j}^{y}\hat{\sigma}_{j+1}^{x}\right)\big].
\end{aligned}
\end{equation}
Here $\Delta_{\Omega=0}$ is an XXZ anisotropy (note that it differs from the anisotropy of the model in Eq.~\eqref{eq_SpinModelXXZ} because we are in the low-drive limit here), and $D$ describes the relative strength of the DM term. In our system the coefficients are (again, maintaining $\Omega = 0$),
\begin{equation}
\label{eq_DMModelCoefficients}
\Delta_{\Omega=0} = \sec(\phi),\>\>\>D = -\tan(\phi).
\end{equation}
The non-zero flux causes this model to deviate away from a conventional Heisenberg model by a set of position-dependent local rotations of the spin variables~\cite{xu2014mottSuperfluidSOCbosons} because we are in a dressed basis, see Eq.~\eqref{eq_BasisRotation}. However, we can map back to a Heisenberg model at the price of changing the initial state. More concretely, the dynamics of
\begin{equation}
    \hat{H}_{\mathrm{Heisen+T}} \>\>\text{ evolving }\ket{\psi_0^{(X)}} = \bigotimes_j \ket{\rightarrow}_j,
\end{equation}
which correspond to the low-drive evolution from the prior section, are equivalent to:
\begin{equation}
\begin{aligned}
    \hat{H}_{\mathrm{Heisen}} &=\frac{J^2}{U}\sum_{j}\vec{\sigma}_{j}\cdot \vec{\sigma}_{j+1} \\
    \text{ evolving }\ket{\psi_0^{\mathrm{Spiral}}} &=e^{\frac{i}{2}\sum_{j} j \phi\hat{\sigma}_{j}^{z}}\bigotimes_{j}\ket{\rightarrow}_j,
\end{aligned}
\end{equation}
with the initial state becoming a spiral in the plane of the DM interaction, rotating by an angle $\phi$ per lattice site under a unitary transformation $\hat{U} = e^{\frac{i}{2}\sum_{j}j \phi \hat{\sigma}_{j}^{z}}$. The mapping is exact for open boundary conditions and $\Omega = 0$. Periodic boundaries can lead to incommensurate mismatch if the flux is not a multiple of $2\pi/L$ (see discussion in Appendix~\ref{app_Boundary}). Hereafter, we refer to the dynamics under $\hat{H}_{\mathrm{Heisen+T}}$ as the gauged frame (working in a dressed basis), and the equivalent dynamics under $\hat{H}_{\mathrm{Heisen}}$ as the un-gauged frame (for which the quantization axis of the Bloch sphere is set by the bare atomic states $g$, $e$ with no site-dependent phases).

\begin{figure*}[htb]
\centering
\includegraphics[width=1\linewidth]{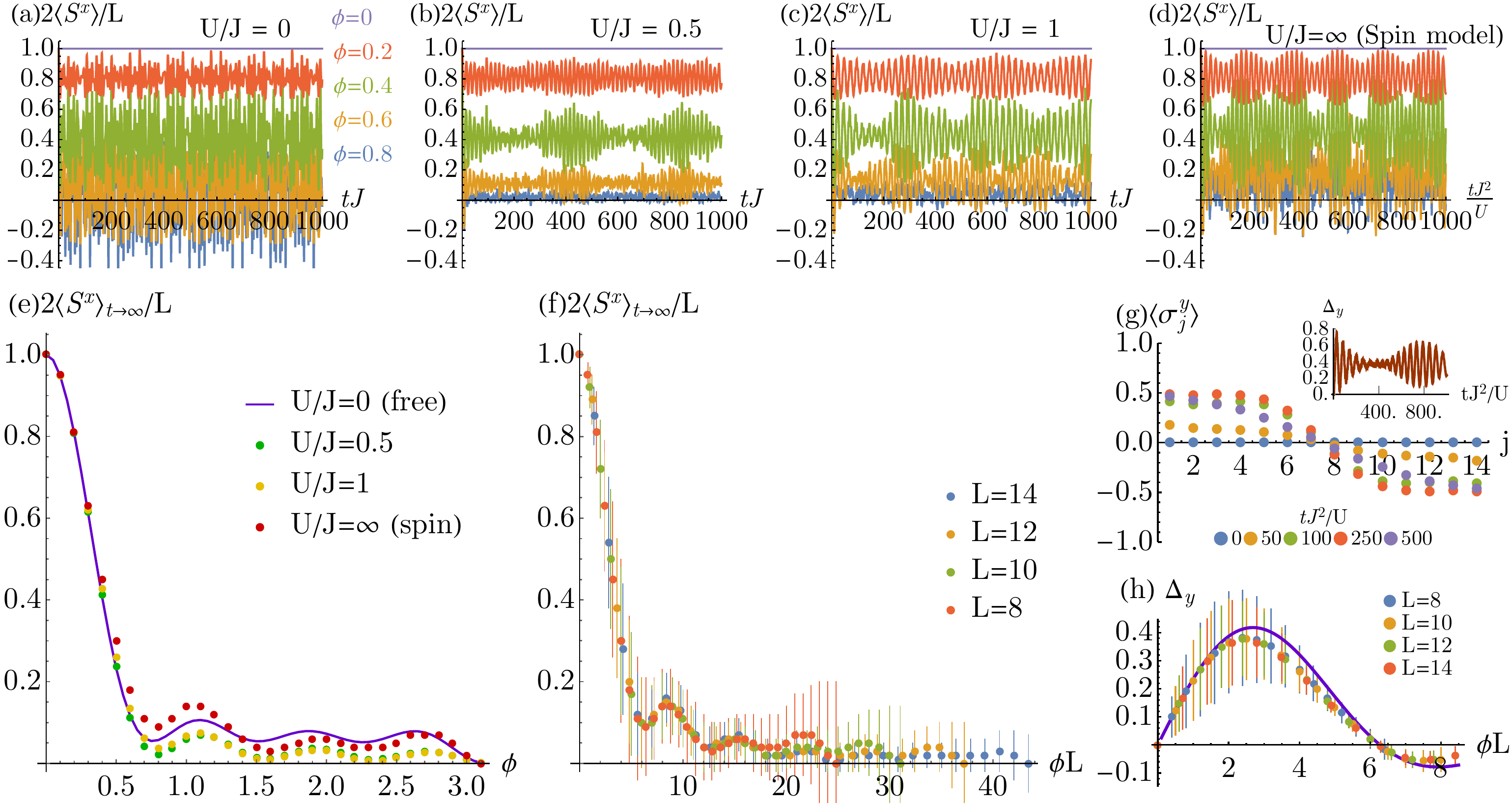}
\caption{(a-d) Time-evolution of the magnetization $\langle \hat{S}^{x}\rangle$ for the Fermi-Hubbard model with $U/J = 0, 0.5, 1$ and $\Omega = 0$ [panels (a-c)] and the Heisenberg+twist spin model ($U/J \to \infty$ limit) [panel (d)], for different values of flux $\phi$. System size is $L=8$. After a short initial decay, the magnetization stabilizes to a nonzero value depending on the flux. (e) Long-time mean magnetization for the different models in panels (a-d). Up to small deviations, the mean is the same for all interaction strengths. (f) Long-time mean magnetization for the spin model as a scaling function of flux times system size $\phi L$ for different system sizes. For smaller flux, the curves fall on top of each other. Larger flux causes them to deviate, as the system is periodic under $\phi \to \phi + 2\pi$ (with no $L$ scaling). Error bars denote 1 standard deviation of the long-time fluctuations about the mean. (g) Site-resolved transverse magnetization $\langle \hat{\sigma}_{j}^{y}\rangle$ for different time snapshots using the spin model ($L=14$). An infinite-time chiral lattice-wide imbalance $\Delta_y$ is established, as plotted in the inset. (h) Plot of infinite-time imbalance, also showing scaling behaviour. Dots are the spin model, while the dark blue line is the $U/J$=0 exact analytic result in the thermodynamic limit $L \to \infty$.}
\label{fig_Magnetization}
\end{figure*}

The spiral structure from non-zero flux causes the system's dynamics to exhibit non-trivial features due to the additional underlying symmetry of the Heisenberg model. As the simplest example of such features, in Fig.~\ref{fig_Magnetization}(a-d) we plot the time-evolution of the collective magnetization $\langle\hat{S}^{x}\rangle$ in the gauged frame (starting from $\ket{\psi_0^{(X)}}$) for different values of flux. We consider both the Fermi-Hubbard model in the appropriate basis for different values of Hubbard repulsion $U/J$ [panels (a-c)] and the spin model $\hat{H}_{\mathrm{Heisen+T}}$ [panel (d)]. We find that in all cases, after an initial decay there is a non-zero mean magnetization that persists to infinite time,
\begin{equation}
\langle\hat{S}^{x}\rangle_{t\to\infty} = \lim_{T \to \infty} \frac{1}{T}\int_{0}^{T} dt \langle\hat{S}^{x}\rangle(t),
\end{equation}
while the other in-plane component $\langle \hat{S}^{y}\rangle$ averages to zero. This dependence of this mean magnetization on the flux is shown in Fig.~\ref{fig_Magnetization}(e). If we had started with a product state along the $\hat{y}$-direction of the Bloch sphere we would have a non-zero mean $\langle \hat{S}^{y}\rangle$ and zero mean $\langle\hat{S}^{x}\rangle$ instead. Interestingly enough, we find that while the amplitude of fluctuations about the mean depends strongly on the Hubbard parameter $U/J$ (recall that the spin model is only valid for $U/J\gg 1$), the mean value remains largely the same independently of $U/J$. This permits us to write an approximate analytic expression for the mean by solving the $U/J = 0$ interactionless case (with open boundaries),
\begin{widetext}
\begin{equation}
\label{eq_MagnetizationAnalytic}
    \frac{2}{L}\langle \hat{S}^{x}\rangle_{t\to \infty} \approx \frac{4}{L(L+1)^2} \sum_{j,j',k=1}^{L}\sin^2\left(\frac{\pi j k}{L+1}\right)\sin^2\left(\frac{\pi j' k}{L+1}\right)\cos\left[\phi(j-j')\right]=\frac{1}{L^2}\frac{\sin^2 \left(\frac{\phi L }{2}\right)}{\sin^2\left(\frac{\phi}{2}\right)}\text{  as $L \to \infty$}.
\end{equation}
\end{widetext}
As an even more intriguing feature, in Fig.~\ref{fig_Magnetization}(f) we find that at small flux $\phi L \lesssim 1$ there is a scaling behaviour as a function of $\phi L$, or flux times total system size, which eventually peels off once $\phi$ gets large enough. The first minimum of this scaling long-time magnetization occurs at $\phi L = 2\pi$, which corresponds to a full-period twisting of a spiral state in the un-gauged frame. We see a non-zero magnetization even at a full period twist because we use open boundary conditions. Periodic boundaries see a similar profile, except with $\langle \hat{S}^{x}\rangle_{t \to \infty}$ falling to zero at $\phi = 2\pi n /L$ for any $n \in \mathbbm{N}$ (see Appendix~\ref{app_Boundary} for further discussion on boundary effects).

An infinite-time non-zero magnetization independent of $U/J$ for this model is surprising. Naively, one would expect a relaxation to zero magnetization, as the system has rotational symmetry in the $\hat{x}$-$\hat{y}$ plane of the Bloch sphere, and there is no obvious conservation law that discriminates $\langle\hat{S}^{x} \rangle$ from $\langle\hat{S}^{y} \rangle$. One could expect this to be a consequence of 1D integrability, since the model maps to a Heisenberg model in the un-gauged frame, but we also find similar effects in equivalent 2D systems (see Sec.~\ref{sec_NonThermalizingFeatures}). One may also consider this to be a fine-tuned regime, but in addition to its persistence for all $U/J$ Hubbard repulsion strengths, we find that this non-zero magnetization is robust to perturbative effects such as harmonic trapping or imperfect filling fraction $N/L < 1$, which only slightly change the outcome (See Appendix~\ref{app_TrapAndFilling}). Furthermore, the scaling behaviour maintaining a non-zero average at $\phi L = 2\pi$ implies that boundary conditions play a non-trivial role even in the thermodynamic limit (this is unique to the interacting spin model, see Appendix~\ref{app_Boundary} for details).

Aside from simple observables like magnetization, the system can also develop long-time lattice-wide chiral spin imbalances. Fig.~\ref{fig_Magnetization}(g) shows the other site-resolved in-plane magnetization $\langle\hat{\sigma}_j^{y}\rangle$ across the lattice for different snapshots of the time-evolution (with open boundaries). While the mean $\langle \hat{S}^{y}\rangle$ is zero, we find that the system establishes a tilt in the spin-projection along the $\hat{y}$ direction, indicating that the DM interaction maintains a non-zero spin current at all times, opposed by the relaxation dynamics of the XXZ model. This tilt can be quantified by,
\begin{equation}
\Delta_y = \frac{1}{L}\left(\sum_{j=1}^{L/2}\langle\hat{\sigma}_{j}^{y}\rangle - \sum_{j=L/2+1}^{L}\langle\hat{\sigma}_{j}^{y}\rangle\right),
\end{equation}
which is plotted in Fig.~\ref{fig_Magnetization}(h) showing the same characteristic scaling behaviour as the long-time magnetization. The system generates an extensive spin imbalance depending on the scaled flux $\phi L$. We can again approximate it using the non-interacting limit $U/J=0$,
\begin{equation}
    \Delta_{y}\approx \frac{2}{L^2} \frac{\sin\left(\frac{\phi L}{2}\right)\sin^2 \left(\frac{\phi L}{4}\right)}{\sin^2 \left(\frac{\phi}{2}\right)} \text{ as $L \to \infty$.}
\end{equation}
This imbalance is again surprising, especially because it manifests as an extensive chiral feature resulting from an initial excitation with non-extensive energy in the thermodynamic limit. Furthermore, for $\phi L \leq 2\pi$ the spin is imbalanced in one direction, whereas for $\phi L$ slightly higher than that the direction is reversed, even though the first $2\pi$ only makes a full-period revolution of the spins and should not set a preferential spin pumping direction.

%%%
\subsection{Symmetry-restricted features}
\label{sec_NonThermalizingFeatures}
%%%
The reason that non-trivial long-time behaviour occurs is because of the underlying exact SU(2) symmetry of the Heisenberg+twist model, together with its associated reduction of available phase space at small values of the flux. While the model we study is in a twisted (gauged) frame, the local basis rotations still preserve the associated conserved quantities, just in a twisted form. With no flux $\phi = 0$ we have a Heisenberg model, which is a critical point of the XXZ model ($\Delta_{\Omega=0} = 1$, $D = 0$). The addition of the DM term causes this critical point to extend into a line $D = \pm\sqrt{\Delta_{\Omega=0}^2-1}$ (with a corresponding branch for the $\Delta_{\Omega=0} = -1$ critical point). For the parameters in Eq.~\eqref{eq_DMModelCoefficients}, the system remains on the critical line at some position determined by the flux. The additional symmetries of the Heisenberg model, while twisted, still cause the Hilbert space to break into symmetry sectors and restrict the number of states the system can relax into. By comparison, a model sitting off the critical line of $\Delta_{\Omega=0} = \sec(\phi), D = -\tan(\phi)$ cannot be mapped to the Heisenberg and will have dynamics that are chaotic even in 1D~\cite{vahedi2016XXZandDMchaos}, causing the magnetization to decay to zero for all $\phi \neq 0$.

Recall that in general, the Heisenberg model conserves both total angular momentum $\vec{S}^2$ [with eigenvalues $S(S+1)$] and angular momentum projection $\hat{S}^{x}$ (with eigenvalues $M$), splitting the Hilbert space into $\vec{S}^2$ shells and $\hat{S}^{x}$ sectors within each shell:
\begin{equation}
\begin{aligned}
&[\hat{H}_{\mathrm{Heisen}}, \vec{S}^{2}]=0,\>\>\>\>S = \frac{L}{2},\> \frac{L}{2}-1,\dots\\
&[\hat{H}_{\mathrm{Heisen}},\hat{S}^{x}] = 0,\>\>\>\>M = S, S-1 \dots -S.
\end{aligned}
\end{equation}
With $\phi \neq 0$, the spiral initial state $\ket{\psi_0^{\mathrm{Spiral}}}$ in the un-gauged frame will be distributed among these symmetry sectors. For sufficiently small flux, only the highest angular momentum shells are populated. These include the Dicke manifold $S =\frac{L}{2}$ and the spin-wave manifold $S=\frac{L}{2}-1$, followed by $S =\frac{L}{2}-2$, etc. The higher angular momentum shells have few states per symmetry sector [$1$ in Dicke, $L-1$ in spin-wave, then $\mathcal{O}(L^2)$, $\mathcal{O}(L^3)$ and so on]. When enough of the initial state population sits in these highest shells, there are insufficient states for the system to relax and an infinite-time magnetization is generated. The equivalence between the Fermi- and spin models can also be understood from this argument; the undriven Fermi-Hubbard model in the un-gauged frame also has SU(2) symmetry regardless of the value of $U/J$, meaning that the lack of relaxation should persist. Spin-insensitive perturbations such as external harmonic trapping or imperfect filling fraction likewise maintain SU(2) symmetry and preserve the magnetization or imbalance. Non-negligible boundary effects in the thermodynamic limit are also sensible, as the structure of the highest angular momentum shells depends strongly on the boundaries (sinusoidal vs plane-wave), causing the relevant populations as a scaling function of $\phi L$ to be different. Note that the non-negligible boundary effects are only maintained in the thermodynamic limit for the interacting spin model, however, as discussed in Appendix~\ref{app_Boundary}. In a sense, the system thermalizes within a restricted set of Hilbert space manifolds that prevent full relaxation in the conventional manner.

\begin{figure*}[htb]
\centering
\includegraphics[width=0.8\linewidth]{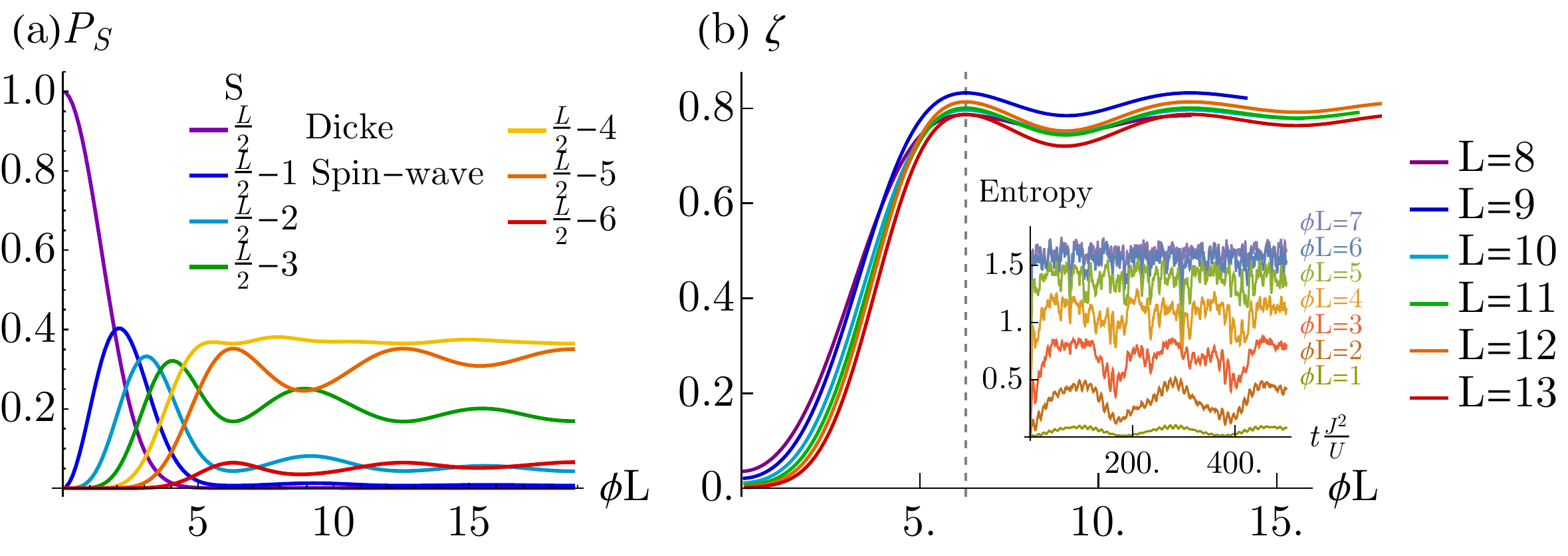}
\caption{(a) Distribution of initial spiral state $\ket{\psi_0^{\mathrm{Spiral}}}$ wavefunction population in different total angular momentum shells $P_S$ of the Heisenberg model $\hat{H}_{\mathrm{Heisen}}$ as a function of scaled flux $\phi L$, for a system of size $L=12$. (c) Weighted average $\zeta$ of shell population times the number of states per symmetry sector of each shell. Note that each shell has a number of symmetry sectors corresponding to the further conserved quantity $S^{x}$. The Dicke manifold has $N_S = 1$, spin-waves $N_S =L-1$, then $\mathcal{O}(L^2)$, $\mathcal{O}(L^3)$ and so on. The dashed line corresponds to $\phi L = 2\pi$. The inset shows the bipartite entanglement entropy dynamics for an $L=10$ system for specific snapshots of flux.}
\label{fig_Thermal}
\end{figure*}

To help quantify the above arguments, in Fig~\ref{fig_Thermal}(a) we plot the overlap of the initial state wavefunction $\ket{\psi_0^{\mathrm{Spiral}}}$ in the un-gauged frame with the different symmetry sectors of the Heisenberg model $\hat{H}_{\mathrm{Heisen}}$,
\begin{equation}
P_{S} = \sum_{M} \sum_{n} |\langle\psi_0^{\mathrm{Spiral}} |\phi_{S, M, n} \rangle|^2,
\end{equation}
where $\ket{\phi_{S, M, n}}$ is the $n$-th eigenstate of $\hat{H}_{\mathrm{Heisen}}$ within the symmetry sector of angular momentum $S$ and projection $M$. At $\phi = 0$, all population sits in the maximally-polarized Dicke state $\bigotimes_j \ket{\rightarrow}_j$. Increasing $\phi$ causes the deeper shells to become populated, increasing the number of states that the system can explore. We give a metric of this property by defining,
\begin{equation}
\label{eq_ThermalMetric}
\zeta = \frac{1}{N_{\mathrm{max}}}\sum_{S} P_S N_{S},
\end{equation}
which is a weighted average of the population in each shell times the number of states per symmetry sector in that shell $N_{S}$, normalized by the number of states in the largest sector $N_{\mathrm{max}}=N_{S=1}$ (for even $L$). The metric $\zeta$ can be understood as a measure of how big a Hilbert space the system can explore. When $\zeta \ll 1$, the wavefunction has most of its weight in symmetry sectors much smaller in dimension $N_{S}$ than the largest-size ones (with size $N_{\mathrm{max}}$), and the system will have trouble relaxing. For $\zeta$ approaching of order one, most the wavefunction weight is in the biggest possible sectors and we can expect a more conventional decay of magnetization/imbalance. Note that $N_{S}$ is not simply the number of states in the $S$-shell, but the number of states per sector of fixed $M$ as well. For example, the Dicke manifold has $L+1$ states in total, but each symmetry sector of $M = -\frac{L}{2}, \dots, \frac{L}{2}$ within it only has 1 state, thus $N_{S = L/2} = 1$. In Fig.~\ref{fig_Thermal}(b) we plot this metric as a function of $\phi L$, finding a characteristic scaling crossover in behaviour. The regime where mean infinite time magnetization falls near zero corresponds to the regime where $\zeta$ saturates to a value near one.

To connect with more conventional metrics, in the inset of Fig~\ref{fig_Thermal}(b) we also plot the dynamics of the bipartite entanglement entropy (partitioning the lattice into left/right halves) for different values of $\phi L$. As $\zeta \approx 1$ is approached, the entanglement entropy saturates at its maximum permitted value based on the system size, while for small flux $\phi L \lesssim 1$ it never reaches that value.

\begin{figure}[htb]
\centering
\includegraphics[width=1\linewidth]{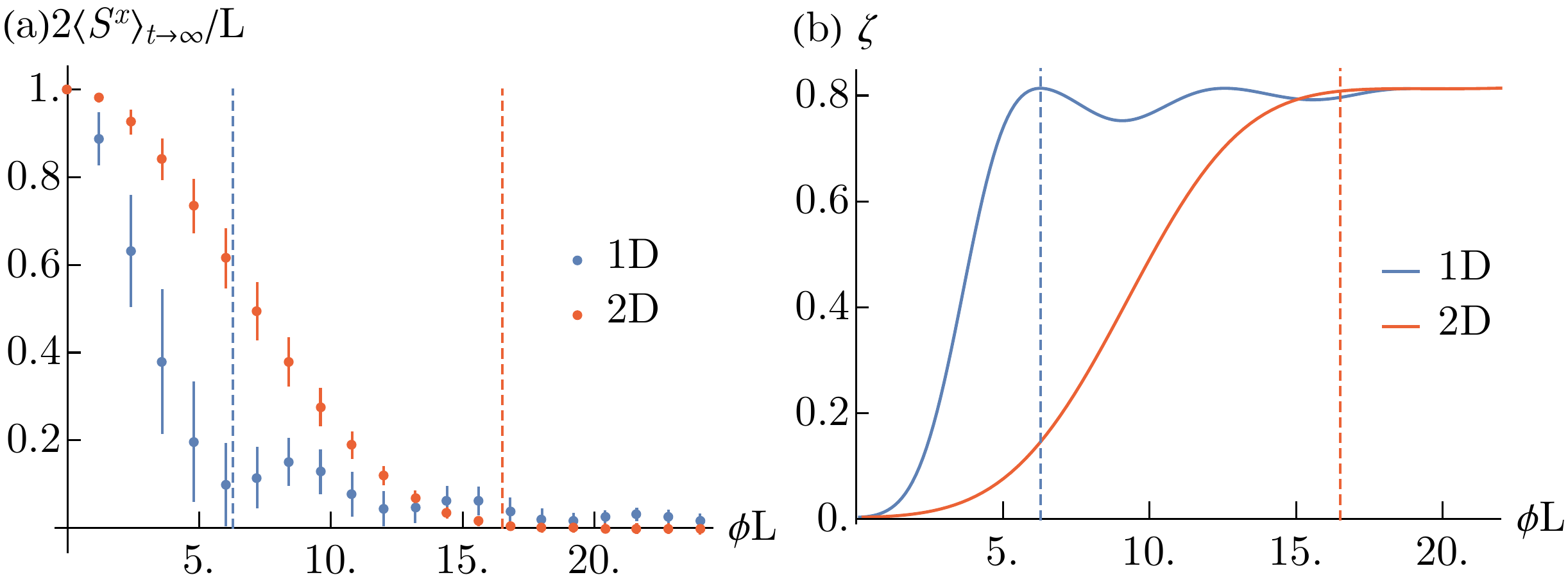}
\caption{(a) Long-time magnetization for the model of Eq.~\eqref{eq_SpinModelDM} and its 2D equivalent as a function of $\phi L$ (in 2D, $L = L_x \times L_y$). System size is $L=12$ in 1D and $(L_x, L_y) = (4,3)$ in 2D, using open boundaries. Error bars indicate one standard deviation of the fluctuations about the mean. Dashed lines estimate the point where the magnetization first reaches a minimum. (b) Hilbert space fragmentation metric from Eq.~\eqref{eq_ThermalMetric} for the same 1D and 2D systems.}
\label{fig_2D}
\end{figure}

The unusual dynamics described above appear reminiscent to other kinds of unusual dynamics associated with integrability, and indeed the 1D nearest neighbor Heisenberg model is integrable. However here we find that integrability is not a necessary (and is in general not a sufficient) ingredient for the observed long-time dynamics. Beyond the mechanism we propose above, which is unrelated to integrability, additional evidence for the unimportance of integrability here is our surprising observation of analogous long-time behaviour in an equivalent 2D Heisenberg model, which is not thought to be integrable. Fig.~\ref{fig_2D}(a) plots the infinite-time magnetization in 2D, using a similar spiral initial state with a 2D structure [i.e. a unitary transformation of the form $e^{\frac{i\phi(i+j)}{2}\hat{\sigma}_{i,j}^{z}}$ for lattice coordinates (i,j)] . While the qualitative profile is changed, we still find non-zero persistent averages, which actually remain higher out to longer values of $\phi L$ (with $L = L_x \times L_y$ for lattice length $L_x$ and width $L_y$). This occurs because 2D systems retain more population in the highest shells for the same flux (since the energy gaps between shells scale with coordination number), and those shells have the same symmetry sector sizes $N_S$ independent of dimension. Fig.~\ref{fig_2D}(b) confirms this prediction by plotting the same metric $\zeta$, showing that it saturates at $\zeta \approx 1$ at the same flux that we see the infinite-time magnetization drop to zero. There have also been studies of similar physics in 3D using approximate numerical methods~\cite{babadi2015spirals3D}, although there persistent infinite-time magnetization was found for $\phi = 0, \pi$.

%%%
\section{Experimental implementations}
\label{sec_Experiment}
%%%

The general system described in Eq.~\eqref{eq_FermiHamiltonian} can be realized in several ways. The most straightforward implementation is with a 3D optical lattice. Dynamics can be restricted to 1D as explored in this work by increasing the transverse lattice depths, e.g. a deep lattice along directions $\hat{y}$, $\hat{z}$ and a shallower depth along a tunneling axis $\hat{x}$ (though still deep enough to maintain $U/J \gg 1$). Spin-orbit coupled driving may be realized through a direct optical transition between long-lived internal states ($g, e$), such as clock states used in conventional atomic clock protocols, which will ensure the coherence times needed for observing spin dynamics. Flux is generated whenever the drive laser wavevector $\vec{k}_{\mathrm{L}}$ has some projection along the tunneling axis $\hat{x}$, and will take the value $\phi = \cos(\theta)(2\pi a)/\lambda_{\mathrm{L}}$, with $a$ the lattice spacing, $\lambda_{\mathrm{L}}$ the driving laser wavelength and $\theta$ the angle between $\vec{k}_{\mathrm{L}}$ and the tunneling axis $\hat{x}$.

The most promising platform candidates are alkaline earth or earth-like atom experiments, as they provide long-lived optically separated internal clock states and magnetic field insensitivity~\cite{kolkowitz2017soc,Bromley2018,Livi2016}. The sample parameters in Table~\ref{table1} were computed for ultracold fermionic alkaline earth $^{87}$Sr in a magic-wavelength lattice, using the long-lived $^{1}S_0$ and $^{3}P_0$ clock states as the internal states $e$ and $g$. Unfavourable $e$-$e$ collisions are mitigated by using a nuclear-spin polarized gas, as wavefunction symmetry forbids any $s$-wave interactions between two $e$ atoms both on the ground vibrational state, while $p$-wave collisions can only act cross-site and are negligible for the lattice depths of $\gtrsim 10 E_r$ that we consider due to the exponential falloff of lattice Wannier functions. Such an implementation is also insensitive to magnetic field fluctuations, as these will only contribute a time-varying single-particle detuning term $\sim \delta(t) \hat{S}^{z}$. Current-generation optical lattice clock  experiments have magnetic field control that permits reduction of $\delta(t)$ to below $0.1$ Hz (when using clock states of a nuclear-spin polarized gas), which is negligible compared to our bare single-particle drive $\Omega$ on the order of kHz.

Alternatively, one may emulate these types of spin physics with nuclear-spin states such as hyperfine states within a given manifold where the spin-orbit coupling is implemented via Raman transitions. The relevant flux in this case will come from the difference in the overall projection of the two Raman beams, e.g. $\phi = [\cos(\theta_1)- \cos(\theta_2)](2\pi a)/\lambda_{\mathrm{L}}$ with $\theta_1$, $\theta_2$ the angles of the beams to the tunneling axis $\hat{x}$. Spontaneous emission effects can be made negligible by detuning further from the intermediate excited state; see Ref.~\cite{Mancini2015} for an example implementation with ultracold $^{87}$Sr using $^{3}P_1$ as an intermediate state.
Depending on the duration of spin dynamics one wishes to emulate, other atomic platforms such as alkali atoms may also prove useful, especially in regimes near resonance $|U| \approx |\Omega|$ where the spin model still holds, but the timescales are faster. There have also been discussions on the use of Lanthanide atoms~\cite{cui2013lanthanide}, which can avoid some of the heating issues typically found in alkali atoms.

\begin{figure}[htb]
\centering
\includegraphics[width=1\linewidth]{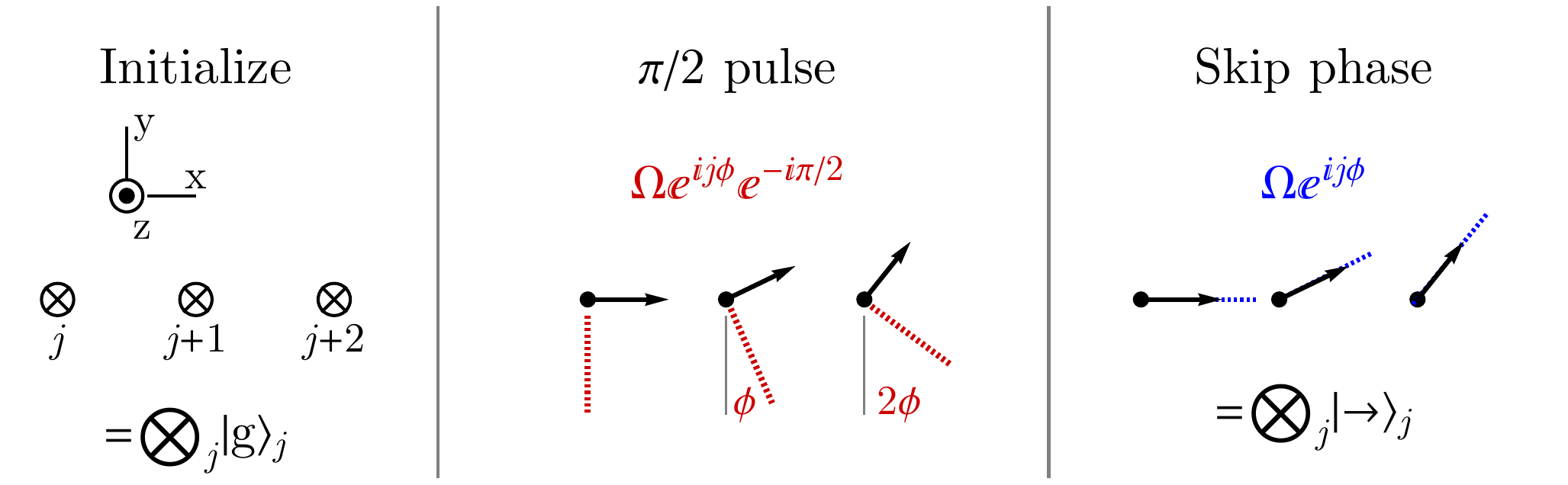}
\caption{Schematic for preparing product states in the dressed eigenbasis of the drive. The system is initialized in a bare atomic product state $\bigotimes_j \ket{g}_j$. The spin-orbit coupled drive is turned on, and a $\pi/2$ pulse is made, preparing a spiral state on the equator of the Bloch sphere (the red lines indicate the drive axes of rotation on each lattice site). The phase of the drive is then skipped by $\pi/2$, shifting the axes to match the current spin direction on each site (blue lines). This results in a product state $\bigotimes_j \ket{\rightarrow}_j$. Measuring $\langle \hat{S}^{x}\rangle$ can also be done by reversing this protocol, then measuring $\sum_j(\langle \hat{n}_{j,e}\rangle-\langle\hat{n}_{j,g}\rangle)$.}
\label{fig_Experiment}
\end{figure}

Preparing the desired product initial states is straightforward. A product state $\ket{\psi_0^{(Z)}}=\bigotimes_j \ket{\uparrow}_j$ in the dressed basis is trivial to prepare, as it is equal to a product state of all atoms in the bare atomic basis up to an overall phase,
\begin{equation}
    \ket{\psi_0^{(Z)}} = \bigotimes_j \ket{e}_j,
\end{equation}
and can thus be initialized with standard optical pumping techniques. Creating a product state $\ket{\psi_0^{(X)}} = \bigotimes_j \ket{\rightarrow}_j$ requires a little more effort, because it is an eigenstate of the drive and cannot be generated with the same drive alone. However, such a state can be prepared by skipping the laser phase. Fig.~\ref{fig_Experiment} shows such a protocol. One initializes all atoms in the bare atomic ground-state $\bigotimes_j \ket{g}_j$, implements a $\pi/2$ pulse with the same laser used for driving, then skips its phase ahead by $\pi/2$. The pulse will create a site-dependent rotation that transforms the state into the desired dressed product state $\bigotimes_j \ket{\rightarrow}_j$,
\begin{equation}
e^{-\frac{i\pi}{4} \sum_j \left(e^{-\frac{i\pi}{2}}e^{i j \phi}\hat{c}_{j,e}^{\dagger}\hat{c}_{j,g} + h.c.\right)}\bigotimes_{j} \ket{g}_j = \bigotimes_{j}\ket{\rightarrow}_j.
\end{equation}
The same techniques may be used for measuring collective observables. Total $\langle \hat{S}^{z}\rangle$ is simply measured from the bare atomic excitation fraction, as $\langle \hat{S}^{z}\rangle = \frac{1}{2}\sum_{j}(\langle \hat{n}_{j,e}\rangle-\langle\hat{n}_{j,g}\rangle)$. Total $\langle \hat{S}^{x}\rangle$ can be measured by reversing the above state preparation protocol; one skips the phase by $-\pi/2$, makes a $\pi/2$ pulse, then measures excitation fraction.

In addition to optical lattices, much of the physics can also be done with optical tweezer arrays placed close enough to allow tunneling. These have seen significant recent development due to their tunability and control~\cite{norcia2018tweezer,endres2016tweezer,saskin2019tweezer,cooper2018tweezer,Browaeys2019,Kaufman2015}, and offer an interesting alternative platform for spin dynamics experiments. If using atoms with long-lived internal clock states, a single interrogating laser can be applied in exactly the same way as for the lattice; one simply needs to control either the angle or tweezer spacing to realize the desired $\phi$. Spin-orbit coupled Raman schemes are likewise straightforward to adapt. As further possibility, higher-dimensional systems are also simple to generate; one makes the lattice shallower along more than one tunneling direction, or uses a 2D tweezer array. Each direction's corresponding flux will be controlled by the projection of the driving laser(s) onto the relevant axis. This permits studies of the same types of spin models in 2D, as well as interplay between interactions of various types along different dimensions (such as for example an isotropic interaction along one dimension and an anisotropic one along another), which can be relevant to emulating real condensed matter materials.

We note that our discussion has focused on systems with ideal filling fraction $N/L = 1$, while real implementations can have holes in the initial loadout. While explicit benchmarking for doped Fermi-Hubbard systems is challenging and often non-representative due to the small numerically-accessible system sizes, we can make an estimate for how good a filling fraction one needs to at least observe the effects of superexchange spin dynamics. A direct comparison of energy scales suggests that the superexchange rates $J_{\parallel}$, $J_{\perp}$ should be greater than or comparable to the bare tunneling rate of holes times the hole fraction, $J(1-N/L)$. Furthermore, for non-zero flux, some of the hole motion is mitigated because part of the tunneling terms acquire a spin-flip in the drive+SOC dressed basis, which is energetically unfavourable when $\Omega \gg J$ [see Appendix~\ref{app_SpinDerivation} for details]. In the case of $\phi = \pi$ the holes are completely locked in place, and in general their tunneling is reduced by a factor of $\cos(\phi/2)$. Altogether this leads to the requirement that $J_{\parallel},J_{\perp} \ll J(1-N/L)\cos(\phi/2)$. Current-generation optical lattice experiments can already reach filling fractions of $(1-N/L) \lesssim 0.1$, for which our typical superexchange rates [Table~\ref{table1}] are comparable to this normalized hole tunneling. Quantum gas microscopes also offer a powerful means of achieving even higher filling where superexchange will dominate the dynamics.

%%%%%
\section{Conclusions}
%%%%%
We have shown that using a single laser drive to induce magnetic flux in a fermionic optical lattice system can realize a wide variety of different spin models across the parameter regime of flux and driving strength. This system is readily implementable in  modern optical lattice or tweezer experiments using highly coherent atomic states. It opens a path to  greatly improve  quantum simulation capabilities using tools already in reach in current experiments. In addition to studying well-understood models such as the Ising or one-axis twisting models, more exotic physics such as lack of relaxation imposed by symmetry constraints can also be explored with this setup.

In addition, the reduction of the Fermi-Hubbard model to well-studied spin models provides a significant advantage for research into the underlying physics. The Fermi-Hubbard has proven challenging to study even in its equilibrium properties. Transforming the system into a simple spin model with well-understood properties allows for a significant coarse-graining in the form of collective observable dynamics, which are more experimentally and theoretically tractable. A spin model mapping of this form shows that the underlying Fermi-Hubbard model is more simple than one would expect if working in the right dressed basis, and conversely allows for easier probes of the system's response to additional ingredients like external fields.

There are many possible future directions to explore on both experimental and theoretical fronts. One may inquire further as to what happens to similar long-time magnetization behaviour in higher-dimensional systems; we find that 2D also exhibits non-zero steady state averages, but there are qualitative differences in the resulting profiles. Boundary effects can be probed, as we find discrepancies between interacting and free models (see Appendix~\ref{app_Boundary}), which could become more complex in higher dimensions. A more in-depth study of the persistent magnetization behaviour's breakdown from external perturbations can also be done. Finally, one can study the steady state spin imbalances in the context of spin transport.

{\bf Acknowledgements.} This material is based in part upon work supported by
the AFOSR grant FA9550-19-1-0275 9 (AMR) and FA9550-20-1-0222 (RN), by the DARPA and ARO grant W911NF-16-1-0576, the ARO single investigator award W911NF-19-1-0210, the NSF PHY1820885, NSF JILA-PFC PHY-1734006 grants, and by NIST.
\bibliography{DMBibliography.bib}
\bibliographystyle{unsrt}

\clearpage

\onecolumngrid
\renewcommand{\thesubsection}{\Alph{subsection}}
\renewcommand{\appendixtocname}{Supplementary material}
\appendixpageoff
\appendixtitleoff
\begin{appendices}

%%%%%%%%%%%%%%%%%%%%%%%
\section{Spin model derivation}
\label{app_SpinDerivation}
\renewcommand{\theequation}{A\arabic{equation}}
\setcounter{equation}{0}
%%%%%%%%%%%%%%%%%%%%%%%

We show here how the spin model we use is derived from the Fermi-Hubbard model. The method is standard second-order Schrieffer-Wolff perturbation theory, as described in e.g. Ref.~\cite{bravyi2011schrieffer}.

The perturbation theory proceeds by first rotating the full Fermi-Hubbard Hamiltonian $\hat{H}$ into the eigenbasis of the drive by defining new fermionic operators,
\begin{equation}
\begin{aligned}
    \hat{b}_{j,+} &= \frac{1}{\sqrt{2}} \left(\hat{c}_{j,e} + e^{i j \phi} \hat{c}_{j,g}\right),\\
    \hat{b}_{j,-} &= \frac{1}{\sqrt{2}} \left(\hat{c}_{j,e} - e^{i j \phi} \hat{c}_{j,g}\right).
\end{aligned}
\end{equation}
The Hamiltonian becomes,
\begin{equation}
\begin{aligned}
    \hat{H}^{'} &= \hat{H}_{J}^{'} +\hat{H}_{U}^{'} +\hat{H}_{\Omega}^{'}\\
    \hat{H}_{J}^{'}&=-\frac{J}{2}\sum_{j}\left[\left(1+e^{-i\phi}\right)\left(\hat{b}_{j,+}^{\dagger}\hat{b}_{j+1,+} + \hat{b}_{j,-}^{\dagger}\hat{b}_{j+1,-}\right)+\left(1-e^{-i\phi}\right)\left(\hat{b}_{j,+}^{\dagger}\hat{b}_{j+1,-} + \hat{b}_{j,-}^{\dagger}\hat{b}_{j+1,+}\right) + h.c.\right],\\
    \hat{H}_{U}^{'}&=U \sum_{j}\hat{n}_{j,+}\hat{n}_{j,-},\\
    \hat{H}_{\Omega}^{'}&=\frac{\Omega}{2}\sum_{j}\left(\hat{n}_{j,+}-\hat{n}_{j,-}\right).
\end{aligned}
\end{equation}
Note that the attenuation of possible hole motion with imperfect filling can be understood from the above equations, as the two terms in $\hat{H}_{J}^{'}$ correspond to spin-conserving and spin-flipping tunneling with amplitudes of $\cos(\phi/2)$ and $\sin(\phi/2)$ respectively (up to an overall phase), with the latter mitigated by the energy penalty imposed by the now-diagonal drive $\hat{H}_{\Omega}^{'}$.

The next step is to split the full Hilbert space of this system into two energetically-separated manifolds. The first is a manifold $\mathbbm{E}_0$ which contains all states with one atom per site only (the effective spin states). The second is a manifold $\mathbbm{E}_{V}$ containing doubly-occupied states (which will not be directly populated during dynamics, but act as virtual intermediate states). In our case, since the tunneling is nearest-neighbour and we are at half filling, it is sufficient to work with the states of a two-site system $j=1,2$ (with two atoms). For such a system, the Hilbert space may be written in terms of Fock states with fermionic occupation numbers indexed as $\ket{n_{1,\uparrow}, n_{1,\downarrow},n_{2,\uparrow},n_{2,\downarrow}}$, and the two manifolds are:
\begin{equation}
\begin{aligned}
    \left(\mathbbm{E}_0\right)_{\mathrm{2 sites}} &= \{\ket{1,0,1,0},\ket{1,0,0,1},\ket{0,1,1,0},\ket{0,1,0,1}\},\\
    \left(\mathbbm{E}_V\right)_{\mathrm{2 sites}} &= \{\ket{1,1,0,0},\ket{0,0,1,1}\}.
\end{aligned}
\end{equation}
General Schrieffer-Wolff theory then prescribes that we split the Hamiltonian into a diagonal portion $\hat{H}_0$ containing the now-diagonal drive and Hubbard interaction, and a block off-diagonal portion $\hat{V}_{\mathrm{od}}$ that couples the manifolds $\mathbbm{E}_0$, $\mathbbm{E}_V$ (with no $\hat{V}_{\mathrm{od}}$ matrix elements within the manifolds themselves). To this end, we define projectors,
\begin{equation}
    \hat{P}_0 = \sum_{a \in \mathbbm{E}_0}\ket{a}\bra{a},\>\>\>\hat{P}_{V} = \sum_{a \in \mathbbm{E}_V}\ket{a}\bra{a}. 
\end{equation}
The Hamiltonian is then separated as,
\begin{equation}
\begin{aligned}
    \hat{H}^{'} &= \hat{H}_{0}+\hat{V}_{\mathrm{od}},\\
    \hat{H}_0 &=\hat{P}_0\left(\hat{H}_{U}^{'} + \hat{H}_{\Omega}^{'}\right)\hat{P}_0 + \hat{P}_{V}\left(\hat{H}_{U}^{'} + \hat{H}_{\Omega}^{'}\right)\hat{P}_{V},\\
    \hat{V}_{\mathrm{od}} &=\hat{P}_0\hat{H}_{J}^{'}\hat{P}_{V} + \hat{P}_{V}\hat{H}_{J}^{'}\hat{P}_{0}.
\end{aligned}
\end{equation}
There could also be off-diagonal but block-diagonal portions $\hat{V}_d$, but in our case these do not occur. This is an intentional result of going to a basis where the drive $\hat{H}_{\Omega}^{'}$ is diagonal. Otherwise, we would have other high-order perturbative effects emerge if the drive is strong. 

We can now write the second-order Schrieffer-Wolff transformation generator as,
\begin{equation}
    \hat{S}_1 = \sum_{a\neq b}\frac{\bra{a}\hat{V}_{od}\ket{b}}{\bra{a}\hat{H}_0\ket{a}-\bra{b}\hat{H}_0\ket{b}}\ket{a}\bra{b},
\end{equation}
where both sums run over \textit{all} states in the Hilbert space, $\ket{a},\ket{b} \in \mathbbm{E}_0\bigcup \mathbbm{E}_V$. The summand should be treated as zero whenever the numerator is zero (i.e. when there is no matrix element between $\ket{a}$, $\ket{b}$), ignoring any zero denominator that can arise from degeneracy. The effective second-order Hamiltonian restricted to the $\mathbbm{E}_0$ manifold is then written as,
\begin{equation}
    \hat{H}_{\mathrm{SE}} = \hat{P}_0\left(\hat{H}_0 + \frac{1}{2}[\hat{S}_1,\hat{V}_{\mathrm{od}}]\right)\hat{P}_0.
\end{equation}
For the specific case of a two-site system $j=1,2$, the resulting Hamiltonian is a $4\times 4$ matrix,
\begin{equation}
    \left(\hat{H}_{\mathrm{SE}}\right)_{\mathrm{2 sites}} = \left(
\begin{array}{cccc}
 \Omega+\frac{J^2 (\cos (\phi )-1)}{U-\Omega } & \frac{i J^2 (2 U-\Omega ) \sin (\phi )}{2 U (U-\Omega )} & -\frac{i J^2 (2 U-\Omega )
   \sin (\phi )}{2 U (U-\Omega )} & -\frac{J^2 U (\cos (\phi )-1)}{U^2-\Omega ^2} \\
 -\frac{i J^2 (2 U-\Omega ) \sin (\phi )}{2 U (U-\Omega )} & -\frac{J^2 (\cos (\phi )+1)}{U} & \frac{J^2 (\cos (\phi )+1)}{U} &
   \frac{i J^2 (2 U+\Omega ) \sin (\phi )}{2 U (U+\Omega )} \\
 \frac{i J^2 (2 U-\Omega ) \sin (\phi )}{2 U (U-\Omega )} & \frac{J^2 (\cos (\phi )+1)}{U} & -\frac{J^2 (\cos (\phi )+1)}{U} &
   -\frac{i J^2 (2 U+\Omega ) \sin (\phi )}{2 U (U+\Omega )} \\
 -\frac{J^2 U (\cos (\phi )-1)}{U^2-\Omega ^2} & -\frac{i J^2 (2 U+\Omega ) \sin (\phi )}{2 U (U+\Omega )} & \frac{i J^2 (2
   U+\Omega ) \sin (\phi )}{2 U (U+\Omega )} & -\Omega+\frac{J^2 (\cos (\phi )-1)}{U+\Omega } \\
\end{array}
\right).
\end{equation}
We can re-write this matrix in terms of Pauli matrices $\hat{\sigma}^{\gamma}$ (with $\gamma \in \{x,y,z\}$),
\begin{equation}
    \left(\hat{H}_{\mathrm{SE}}\right)_{\mathrm{2 sites}} = J_{\perp}\hat{\sigma}_{1}^{x}\hat{\sigma}_{2}^{x} + J_{\parallel}\left(\hat{\sigma}_{1}^{y}\hat{\sigma}_{2}^{y}+\hat{\sigma}_{1}^{z}\hat{\sigma}_{2}^{z}\right) + J_{\mathrm{DM}}\left(\hat{\sigma}_{1}^{z}\hat{\sigma}_{2}^{y} - \hat{\sigma}_{1}^{y}\hat{\sigma}_{2}^{z}\right) +\frac{J^2 \Omega  \sin (\phi )}{2(U^2-\Omega ^2)} \left(\hat{\sigma}_{1}^{y} - \hat{\sigma}_{2}^{y}\right)+ \left[\frac{\Omega}{2} -\frac{J^2 \Omega \sin^2 \left(\frac{\phi}{2}\right)}{U^2 - \Omega^2}\right] \left(\hat{\sigma}_{1}^{z} + \hat{\sigma}_{2}^{z}\right),
\end{equation}
where $\hat{\sigma}_{1}^{\gamma} = \hat{\sigma}^{\gamma}\otimes \mathbbm{1}$, $\hat{\sigma}_{2}^{\gamma} = \mathbbm{1} \otimes \hat{\sigma}^{\gamma}$, and the coefficients are the same as the ones in main text Eq.~\eqref{eq_SpinModelCoefficients}.

This is the interaction for two lattice sites. For a full chain of $L > 2$, we simply interpolate it,
\begin{equation}
    \hat{\sigma}_{1}^{\gamma} \to \hat{\sigma}_{j}^{\gamma}, \>\>\> \hat{\sigma}_{2}^{\gamma} \to \hat{\sigma}_{j+1}^{\gamma},
\end{equation}
and sum over $j$. Note that the term with $(\hat{\sigma}_{1}^{y} - \hat{\sigma}_{2}^{y}) \to (\hat{\sigma}_{j}^{y} - \hat{\sigma}_{j+1}^{y})$ will end up cancelling out aside from boundary terms, since for any given lattice site $j$ in the bulk it gets added by the chain link to the left, and subtracted by the chain link to the right. We neglect it altogether. For the $(\hat{\sigma}_{1}^{z} + \hat{\sigma}_{2}^{z})$ term, we double the magnitude of the $J^2$-proportional piece of the coefficient (because it gets added twice in 1D, once by each chain link touching the lattice site), but keep the bare $\Omega/2$ as-is (because it comes from a zeroth-order Hamiltonian and not a superexchange chain link). Overall, we end up with,
\begin{equation}
\begin{aligned}
\hat{H}_{\mathrm{SE}}= J_{\perp}\sum_{j}\hat{\sigma}_{j}^{x}\hat{\sigma}_{j+1}^{x}+ J_{\parallel}\sum_{j}\left(\hat{\sigma}_{j}^{y}\hat{\sigma}_{j+1}^{y}+\hat{\sigma}_{j}^{z}\hat{\sigma}_{j+1}^{z}\right)+J_{\mathrm{DM}} \sum_{j}\left(\hat{\sigma}_{j}^{z}\hat{\sigma}_{j+1}^{y}-\hat{\sigma}_{j}^{y}\hat{\sigma}_{j+1}^{z}\right)+J_{\Omega}\sum_{j}\hat{\sigma}_{j}^{z}.
\end{aligned}
\end{equation}

Finally, while we derived this model in the basis where the drive is diagonal, it is helpful to rotate back to a more conventional form where the drive is of the $\hat{\sigma}^{x}$ type instead. We thus make the rotation,
\begin{equation}
    \hat{\sigma}_{j}^{x} \to -\hat{\sigma}_{j}^{z},\>\>\>\hat{\sigma}_{j}^{y}\to \hat{\sigma}_{j}^{y},\>\>\>\hat{\sigma}_{j}^{z} \to \hat{\sigma}_{j}^{x},
\end{equation}
which corresponds to a fermionic basis of,
\begin{equation}
\begin{aligned}
    \hat{a}_{j,\uparrow} &= \frac{1}{\sqrt{2}} \left(\hat{b}_{j,+}+\hat{b}_{j,-}\right) = \hat{c}_{j,e},\\
    \hat{a}_{j,\downarrow} &= \frac{1}{\sqrt{2}} \left(\hat{b}_{j,+}-\hat{b}_{j,-}\right) = e^{i j \phi}\hat{c}_{j,g},\\
\end{aligned}
\end{equation}
leading to the Hamiltonian of Eq.~\eqref{eq_SpinModel} in the main text.

%%%%%%%%%%%%%%%%%%%%%%%
\section{Spin model dynamic regimes}
\label{app_SpinDynamics}
\renewcommand{\theequation}{B\arabic{equation}}
\setcounter{equation}{0}
%%%%%%%%%%%%%%%%%%%%%%%

While the XXZ and OAT models in the main text are defined in the large drive regime $\Omega \gg J_{\parallel},J_{\perp},J_{\mathrm{DM}}$, analogous versions can also be written for the no-drive case $\Omega = 0$. For the XXZ, we can simply write the full spin model $\hat{H}_{\mathrm{SE}}$ and ignore the DM term,
\begin{equation}
    \hat{H}_{\mathrm{XXZ}}|_{\Omega = 0} = \frac{J^2}{U}\cos(\phi)\sum_{j} \left(\hat{\sigma}_{j}^{x}\hat{\sigma}_{j+1}^{x} + \hat{\sigma}_{j}^{y}\hat{\sigma}_{j+1}^{y}\right) + \frac{J^2}{U}\sum_{j} \hat{\sigma}_{j}^{z} \hat{\sigma}_{j+1}^{z}.
\end{equation}
This can be valid if the DM's prefactor $\sim \sin(\phi)$ vanishes for flux close to 0 or $\pi$. For the OAT, we project the above model into the Dicke manifold,
\begin{equation}
\begin{aligned}
\hat{H}_{\mathrm{OAT}}|_{\Omega=0}&=\hat{P}_{\mathrm{Dicke}}\hat{H}_{\mathrm{XXZ}}|_{\Omega=0}\hat{P}_{\mathrm{Dicke}}\\
&=\frac{4J^2}{U}\frac{\cos(\phi)}{L-1}\vec{S}\cdot \vec{S} + \frac{4J^2}{U}\frac{1-\cos(\phi)}{L-1}\hat{S}^{z}\hat{S}^{z}.
\end{aligned}
\end{equation}

For the dynamical regime diagram in main text Fig.~\ref{fig_PhaseDiagram}, the color scheme is determined as follows. The error metric $\Delta_{\mathrm{Spin}}$ is obtained for each of the spin models $\mathrm{Spin}\in \{\text{Ising, XY, OAT, XXZ, SE}\}$, and truncated to a chosen maximum error $\eta = 0.25$ via $\Delta_{\mathrm{Spin}}\to \min (\Delta_{\mathrm{Spin}},\eta)$, which is the threshold at which the model's color will vanish. Note that the last model "SE" is not the Heisen+T model, but the full spin model, which we use to determine the relevance of the Heisen+T by comparing the SE error to an XXZ error, i.e. checking whether the DM term is relevant. Each point in parameter space is assigned an RGB color coordinate $(r,g,b)$ with color values $r,g,b \in [0,1]$. The different spin models are likewise assigned colors, with red (1,0,0) for Ising, green (0,1,0) for XY, yellow (1,1,0) for OAT, white (1,1,1) for XXZ and blue (0,0,1) for Heisen+T (which, again, will be determined by the difference in error of the SE and XXZ models). The color coordinates are then determined via,
\begin{equation}
\begin{aligned}
    r &= 1 - \left(1-\frac{\Delta_{\mathrm{XY}}}{\eta}\right) - \frac{\Delta_{\mathrm{XXZ}}-\Delta_{\mathrm{SE}}}{\eta},\\
    g &= 1 - \left(1-\frac{\Delta_{\mathrm{Ising}}}{\eta}\right) - \frac{\Delta_{\mathrm{XXZ}}-\Delta_{\mathrm{SE}}}{\eta},\\
    b &= 1 - \left(1-\frac{\Delta_{\mathrm{Ising}}}{\eta}\right) - \left(1-\frac{\Delta_{\mathrm{XY}}}{\eta}\right) - \left(1-\frac{\Delta_{\mathrm{OAT}}}{\eta}\right).
\end{aligned}
\end{equation}
We essentially subtract $(1-\Delta_{\mathrm{Spin}}/\eta)$ from every color that does not use the corresponding spin model. For example, since the Ising is red, we subtract one minus its (scaled) error metric from green and blue. The only exception is the Heisen+T model, for which we instead subtract the difference in scaled error between the XXZ and SE models $(\Delta_{\mathrm{XXZ}}-\Delta_{\mathrm{SE}})/\eta$, as error in the XXZ but not in the SE means that the DM term is relevant. Note that for the row of points with $\Omega = 0$ (which are the only points where the DM plays a non-trivial role due to the scale of the plot), we use the undriven versions of the XXZ and OAT models, $\hat{H}_{\mathrm{XXZ}}|_{\Omega=0}$ and $\hat{H}_{\mathrm{OAT}}|_{\Omega=0}$.  After computing the colors, we scale them by the overall quality of the full spin model, $\nu \to \nu (1-\Delta_{\mathrm{SE}}/\eta)$ for $\nu \in \{r,g,b\}$, to capture the resonances where no spin model description works.

%%%%%%%%%%%%%%%%%%%%%%%
\section{Filling fraction and harmonic trapping robustness}
\label{app_TrapAndFilling}
\renewcommand{\theequation}{C\arabic{equation}}
\setcounter{equation}{0}
%%%%%%%%%%%%%%%%%%%%%%%

\begin{figure}[htb]
\centering
\includegraphics[width=0.8\linewidth]{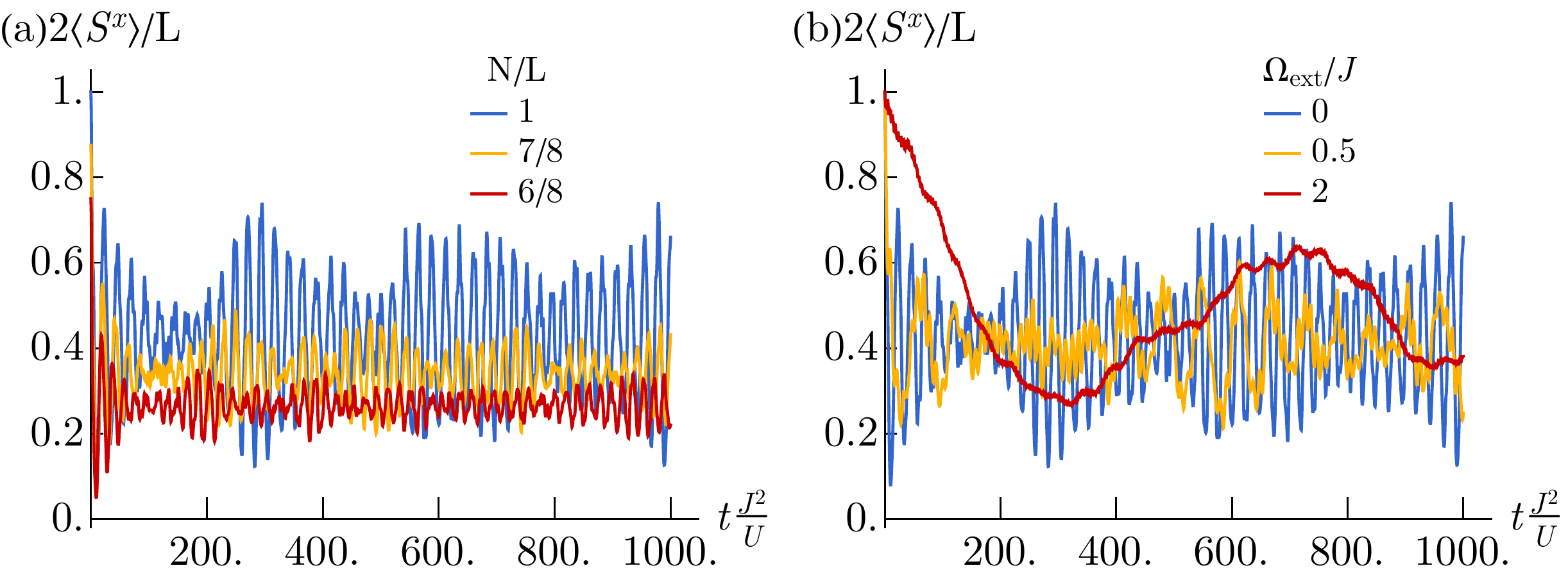}
\caption{(a) Magnetization dynamics of the undriven ($\Omega=0$) system with imperfect filling fraction, modeled using the Fermi-Hubbard model in the dressed basis $\{\uparrow,\downarrow\}$. Lattice size is set to $L =8$ with open boundaries. Fillings of 7/8 and 6/8 are implemented by putting a hole at site $j=4$, and at sites $j=4,6$ respectively (as sample representative evolutions). Hubbard repulsion is set to $U/J = 1$, flux to $\phi = 0.4$. (b) Magnetization in the presence of external harmonic trapping of different strengths $\Omega_{\mathrm{ext}}/J$. Filling is kept ideal at $N/L = 1$, other parameters are the same as in panel (a).}
\label{fig_TrapAndFilling}
\end{figure}

The persistent magnetization/imbalance properties we see are robust to typical optical lattice imperfections like reduced filling fraction, or harmonic trapping coming from lattice beam curvature. The filling fraction can be modeled by directly replacing some of the lattice site atoms with holes in the product initial-state. Fig.~\ref{fig_TrapAndFilling}(a) shows the magnetization for different filling fractions with a small system of $L=8$ using the Fermi-Hubbard model. We find that while there is a reduction of $\langle \hat{S}^{x}\rangle_{t \to \infty}$, the non-zero infinite-time average persists, in agreement with our symmetry-based arguments.

A harmonic trap can be modeled by adding an additional Hamiltonian term,
\begin{equation}
    \hat{H}_{\mathrm{Trap}} = \Omega_{\mathrm{ext}}\sum_{j} \left(j-j_0\right)^2 \left(\hat{n}_{j,e} + \hat{n}_{j,g}\right),
\end{equation}
with $\Omega_{\mathrm{ext}}$ the trap energy. Fig.~\ref{fig_TrapAndFilling}(b) shows the magnetization with the trap present, showing that the long-time averages likewise persist.

%%%%%%%%%%%%%%%%%%%%%%%
\section{Boundary conditions and interaction effects for long-time magnetization}
\label{app_Boundary}
\renewcommand{\theequation}{D\arabic{equation}}
\setcounter{equation}{0}
%%%%%%%%%%%%%%%%%%%%%%%

As discussed in the main text, the dynamics of the Heisenberg+twist model $\hat{H}_{\mathrm{Heisen+T}}$ evolving from a product state $\ket{\psi_0^{(X)}} = \bigotimes_{j}\ket{\rightarrow}_j$ lead to a non-zero infinite-time magnetization $\langle \hat{S}^{x}\rangle_{t \to \infty}$. The analogous dynamics of the Fermi-Hubbard model in the same basis lead to similar infinite-time magnetization values. However, there are some deviations depending on the boundary conditions and system size. 

\begin{figure}[htb]
\centering
\includegraphics[width=0.8\linewidth]{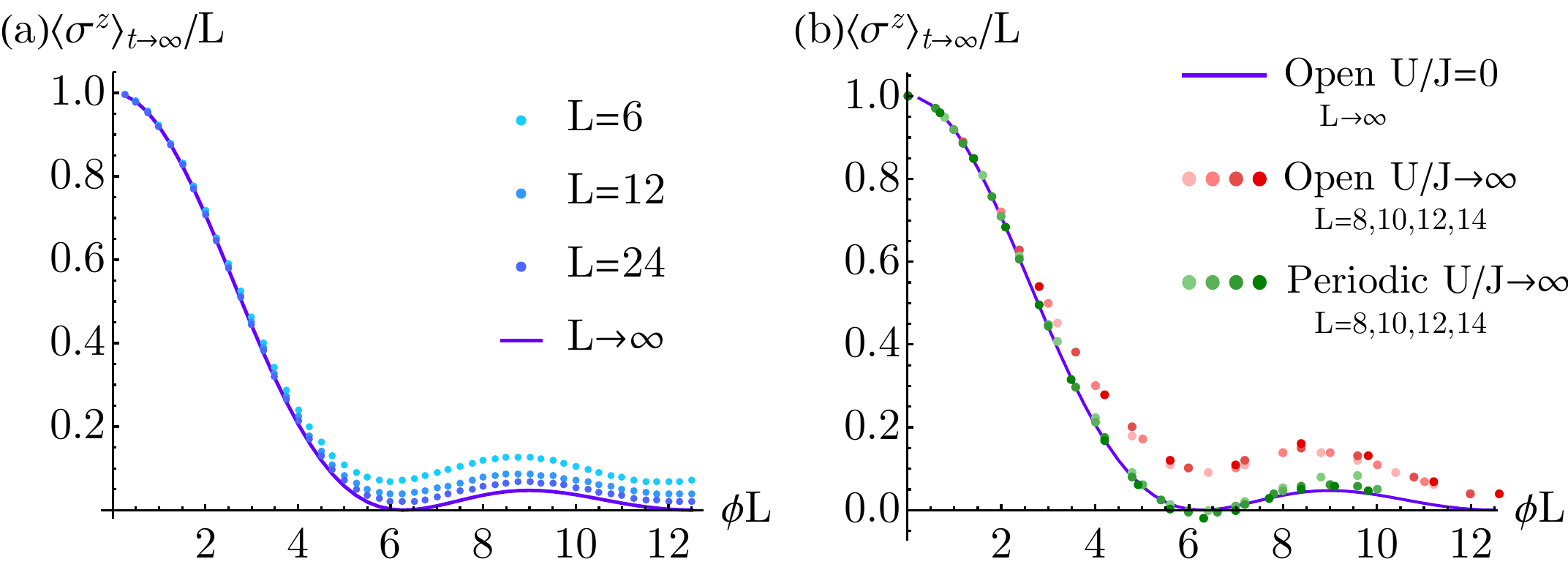}
\caption{(a) Exact long-time magnetization of the non-interacting $U/J = 0$ Fermi-Hubbard model in the gauged frame for different system sizes, using open boundary conditions. The $L \to \infty$ line is the asymptotic formula from main text Eq.~\eqref{eq_MagnetizationAnalytic}. (b) Long-time magnetization comparison between the free and interacting models. The blue line is non-interacting open-boundary $U/J=0$ Fermi-Hubbard model in the thermodynamic limit [same as in panel (a)]. The red dots are the spin model ($U/J \to \infty$ limit) with open boundary conditions and fixed system sizes. The green dots are the same spin model, using periodic boundary conditions.}
\label{fig_BoundarySpinFermi}
\end{figure}

We first consider the infinite-time magnetization $\langle \hat{S}^{x}\rangle_{t \to \infty}$ for the Fermi-Hubbard model with open boundaries in the non-interacting limit $U/J = 0$. The exact result in this regime is given in Eq.~\eqref{eq_MagnetizationAnalytic}, which we re-write here for clarity:
\begin{equation}
    \frac{2}{L}\langle \hat{S}^{x}\rangle_{t\to \infty} \approx \frac{4}{L(L+1)^2} \sum_{j,j',k=1}^{L}\sin^2\left(\frac{\pi j k}{L+1}\right)\sin^2\left(\frac{\pi j' k}{L+1}\right)\cos\left[\phi(j-j')\right]=\frac{1}{L^2}\frac{\sin^2 \left(\frac{\phi L }{2}\right)}{\sin^2\left(\frac{\phi}{2}\right)}\text{  as $L \to \infty$}.
\end{equation}
In Fig.~\ref{fig_BoundarySpinFermi}(a) we plot this value for a few increasing finite values of $L$, as well as the asymptotic form in the thermodynamic limit. It is evident that while there is a non-zero mean at the commensurate flux values $\phi L = 2\pi, 4\pi$, etc. due to the boundaries, this mean vanishes in the thermodynamic limit, which should be expected from a non-interacting model.

The situation becomes more complex when considering the $U/J \to \infty$ limit (i.e. the spin model). Fig.~\ref{fig_BoundarySpinFermi}(b) plots the infinite-time magnetization for the spin model for a few system sizes as a scaling function of $\phi L$ for both open and periodic boundary conditions. We find that there is a clear difference between the two that persists in the thermodynamic limit due to the scaling behaviour. The periodic boundary case sees the magnetization drop to zero at $\phi L = 2\pi, 4\pi$, matching the non-interacting model's thermodynamic limit behaviour (as shown by the blue line in comparison). However, the spin model with open boundaries maintains a non-zero magnetization difference even in the thermodynamic limit, which can be viewed as the difference between the green and red points in the figure. We thus see a purely spin-interaction contribution on top of the persistent magnetization behaviour. The SU(2) symmetry creates a lack of conventional relaxation, leading to non-zero values. However, open boundaries cause the interacting system to have non-zero values even at commensurate $\phi L = 2 \pi, 4\pi$, where the thermodynamic limit should expect to see $\langle \hat{S}^{x}\rangle_{t\to \infty} = 0$, since we have a full-period spiral that should not favor one $\hat{x}$-direction over another.

As a final note, we point out that the spin models used in the above plot are in the un-gauged frame, meaning that we evolve under a Heisenberg model $\hat{H}_{\mathrm{Heisen}}$ from a spiral initial state. For open boundary conditions there is no difference between this and evolving under $\hat{H}_{\mathrm{Heisen+T}}$ with a product initial state, as discussed in the main text. For periodic boundaries, however, the two models do not map correctly to one another unless the flux is commensurate ($\phi L = 2\pi, 4\pi, \dots$), and so we use the un-gauged frame. If we were to evolve a product state with a periodic-boundary version of $\hat{H}_{\mathrm{Heisen+T}}$, we would find that the magnetization decays to zero for all non-zero flux.

\end{appendices}

\end{document}